\documentclass[aip,pop,preprint,a4paper]{revtex4-1}
\usepackage{cancel}
\usepackage{amsmath}
\usepackage{amssymb}
\usepackage{amsthm}
\usepackage{xcolor}

\def\ov#1{\overline{#1}}

\def\vb#1{\mbox{\boldmath$#1$}}
\def\pd#1#2{\frac{\partial #1}{\partial #2}}

\def\wh#1{\widehat{#1}}
\def\bdot{\,\vb{\cdot}\,}
\def\btimes{\,\vb{\times}\,}

\def\bhat{\wh{{\sf b}}}
\def\cal#1{\mathcal{#1}}

\def\exd{{\sf d}}
\def\bhat{\wh{{\sf b}}}

\begin{document}

\title{Guiding-center dynamics in a screw-pinch magnetic field}
\author{A. J. Brizard}
\affiliation{Department of Physics, Saint Michael's College, Colchester, VT 05439, USA \\ {\rm Corresponding author: abrizard@smcvt.edu}}
\date{\today}

\begin{abstract}
    The guiding-center dynamics of charged particles moving in a doubly-symmetric screw-pinch magnetic field is investigated. In particular, we verify that Kruskal's adiabatic-invariant series expansion of the radial action integral associated with the reduced full-orbit radial motion matches the perturbation expansion of the magnetic-moment gyroaction up to first order in magnetic-field non-uniformity. Because the radial action integral is an exact invariant of the full-orbit dynamics, the magnetic moment is therefore represented as non-perturbative integral expression, which can be used to test the validity of the guiding-center approximation.
\end{abstract}

\maketitle

\section{Introduction}

The foundations for the magnetic confinement of charged particles rely on the existence of a hierarchy of adiabatic
invariants\cite{Northrop_Teller_1960,Northrop_1963,Northrop_Liu_Kruskal_1966,Tao_Chan_Brizard_2007,Cary_2009} associated with the three orbital time scales $(T_{\rm g} \ll T_{\rm b} \ll T_{\rm d})$ of the charged-particle motion, which is quasi-periodic on each these time scales. The shortest orbital time scale involves the periodic gyromotion (g) about a single magnetic field line, which is associated with the adiabatic invariance of the gyro-action $J_{\rm g}$. Next, the intermediate orbital time scale involves the bounce (b) motion along a magnetic field line, which is associated with the adiabatic invariance of the bounce action $J_{\rm b}$ when the parallel motion is periodic. Finally, the longest orbital time scale involves the drift (d) motion across magnetic field lines, which is associated with the adiabatic invariance of the  drift action $J_{\rm d}$ when the drift motion is periodic. 

One of the hallmarks of adiabatic invariance is that, while the adiabatic invariant $J_{k}$ is not an exact invariant $dJ_{k}/dt \neq 0$ on the time scale $T_{k}$, the orbital-averaged time derivative yields the identity\cite{Northrop_Teller_1960} $\langle dJ_{k}/dt\rangle_{k} \equiv 0$ when averaged over the time scale $T_{k}$. We note that the adiabatic invariance on a longer orbital time scale requires adiabatic invariance on shorter orbital time scales, e.g., bounce adiabatic invariance requires gyromotion adiabatic invariance (see Fig.~1 of Northrop {\it et al.}\cite{Northrop_Liu_Kruskal_1966}). In addition, the existence of symmetries in the confining magnetic field can alter the adiabatic-invariant hierarchy. For example, when we consider the motion of charged particles (with mass $m$ and charge $e$) in an axisymmetric dipole magnetic field\cite{Brizard_Markowski_2022} ${\bf B} \equiv \nabla\psi\btimes\nabla\varphi$, the toroidal canonical momentum $P_{\varphi} = (e/c)\,\psi + m\,|\partial{\bf x}/\partial\varphi|^{2}\,d\varphi/dt$ is an exact invariant, which can be used as the exact drift action. The analysis of the bounce-center dynamics in an axisymmetric dipole magnetic field can also be done perturbatively\cite{Duthoit_Brizard_2010} when the bounce point along the magnetic-field line is at a shorter distance from the equatorial point than the equatorial radius.

\subsection{Adiabatic invariance of the magnetic moment}

The adiabatic motion of charged particles in a non-uniform magnetic field\cite{Northrop_1963} is associated with the adiabatic invariance of the magnetic moment\cite{Cary_2009}, which can be defined as an action integral\cite{Kruskal_1962} over the periodic gyroangle $\zeta$:
\begin{equation}
    \mu_{\rm gc} \;\equiv\; \frac{\epsilon}{2\pi}\int_{0}^{2\pi}{\bf P}_{\rm gc}\bdot\pd{\vb{\rho}_{\rm gc}}{\zeta}\;d\zeta.
    \label{eq:mugc_integral}
\end{equation}
Here, the dimensionless parameter $\epsilon$ appears as a result of the renormalization of the particle's mass $m \rightarrow \epsilon\,m$ as a way to introduce the dimensional mass-over-charge ratio\cite{Northrop_1963} $m/e$ as an expansion parameter, while the guiding-center canonical momentum ${\bf P}_{\rm gc} \equiv {\sf T}_{\rm gc}^{-1}{\bf P}$ is defined as the guiding-center push-forward of the canonical momentum ${\bf P} \equiv (e/c)\,{\bf A}({\bf x}) + \epsilon\,m{\bf v}$, and $\epsilon\,\vb{\rho}_{\rm gc} \equiv {\sf T}_{\rm gc}^{-1}{\bf x} -{\bf X}_{\rm gc}$ is the guiding-center gyroradius, which is defined as the difference between the guiding-center push-forward of the particle position ${\bf x}$ and the guding-center position ${\bf X}_{\rm gc}$. In addition, the Lie-transform guiding-center push-forward operator\cite{Littlejohn_1982} ${\sf T}_{\rm gc}^{-1}$ involved in Eq.~(\ref{eq:mugc_integral}) is defined for an arbitrary phase-space function $f$ by the series expansion in powers of $\epsilon$:
\begin{equation}
    {\sf T}_{\rm gc}^{-1}f \;\equiv\; \cdots {\sf T}_{3}^{-1}{\sf T}_{2}^{-1}{\sf T}_{1}^{-1}f \;=\; f \;-\; \epsilon\,{\cal L}_{1}f \;-\; \epsilon^{2} \left( {\cal L}_{2}f \;-\; \frac{1}{2}\,{\cal L}_{1}^{2}f \right) + \cdots,
\end{equation}
where the Lie-transform operator ${\sf T}_{n}^{-1} \equiv \exp(-\epsilon^{n}{\cal L}_{n})$ is defined in terms of the $n$th-order Lie derivative ${\cal L}_{n} \equiv {\sf G}_{n}\cdot\exd$ generated by the vector field ${\sf G}_{n}$, whose purpose is to remove the gyroangle dependence from the guiding-center dynamics. The expansion of the action integral (\ref{eq:mugc_integral}) in powers of $\epsilon$ yields the so-called Kruskal asymptotic series\cite{Kruskal_1962} for the guiding-center magnetic-moment adiabatic invariant.

While the integral definition (\ref{eq:mugc_integral}) for the guiding-center magnetic moment $\mu_{\rm gc}$ yields the correct expression for the lowest-order magnetic moment $\mu_{0}({\bf x},{\bf v}) \equiv \epsilon^{2}\,m|{\bf v}_{\bot}|^{2}/(2\,eB({\bf x})/m)$, which is written here with units of action, it is rarely used to obtain higher-order terms because it involves multiple perturbation expansions in powers of $\epsilon$ before performing the gyroangle averages needed to obtain an explicit expression at a desired perturbation order. Instead, by using the standard guiding-center perturbation theory\cite{Littlejohn_1983,Cary_2009} associated with the phase-space transformation from particle coordinates $({\bf x},{\bf p} = m{\bf v})$ to guiding-center coordinates $({\bf X}_{\rm gc},P_{\|{\rm gc}},\mu_{\rm gc},\zeta_{\rm gc})$, the guiding-center magnetic moment is defined as an explicit asymptotic expansion in powers of $\epsilon$:
\begin{equation}
    \mu_{\rm gc}({\bf x},{\bf v}) \;\equiv\; \mu_{0} \;+\; \epsilon\,\mu_{1} \;+\; \epsilon^{2}\,\mu_{2} \;+\; \cdots \;=\; \mu_{0} \;+\; \epsilon\,G_{1}^{\mu} \;+\; \epsilon^{2} \left( G_{2}^{\mu} \;+\; \frac{1}{2}\,{\sf G}_{1}\cdot\exd G_{1}^{\mu} \right) \;+\; \cdots, 
    \label{eq:mugc_series}
\end{equation}
where the first-order correction $\mu_{1} \equiv G_{1}^{\mu}$ is associated with first-order magnetic-field non-uniformity, while the terms $\mu_{n}$ ($n \geq 2)$ involve higher-order corrections\cite{burby_automation_2013,tronko_lagrangian_2015}. Unfortunately, there are concerns about the validity of the asymptotic expansion (\ref{eq:mugc_series}) when the dimensionless parameter $\epsilon$ is not small, either because of steep gradients\cite{Brizard_2017} or when considering energetic particles\cite{j_w_burby_nonperturbative_2025}. It should be pointed out, however, that the comparison of full-orbit particle trajectories with guiding-center orbits depends on the faithfulness of the guiding-center equations, as was noted by Belova {\it et al.} \cite{Belova_2003} and Brizard and Hodgeman \cite{Brizard_2023}. It has been shown by Brizard and Hodgeman \cite{Brizard_2023}, for example, that the traditional guiding-center formalism \cite{Littlejohn_1983} can be improved by retaining corrections associated with guiding-center polarization \cite{tronko_lagrangian_2015}, which were not included in the guiding-center studies cited by Burby {\it et al.} \cite{j_w_burby_nonperturbative_2025}.

A more important comment concerns the conundrum associated with the comparison the integral expression \eqref{eq:mugc_integral} with the local expression \eqref{eq:mugc_series}. On the one hand,  the integral expression \eqref{eq:mugc_integral} is explicitly independent of the gyroangle while, on the other hand,  the local expression \eqref{eq:mugc_series} is explicitly dependent on the gyroangle. In the original work on non-perturbation magnetic-moment invariance\cite{j_w_burby_nonperturbative_2025}, where the magnetic field ${\bf B} = B(x,y)\,\wh{\sf z}$ has a single spatial symmetry axis, a Poincar\'{e} approach is adopted where the local expression \eqref{eq:mugc_series} is restricted to the Poincar\'{e} section defined at the gyroangle $\zeta = 0$ [see Eq.~(4)\cite{j_w_burby_nonperturbative_2025}], while the integral expression \eqref{eq:mugc_integral} is estimated by sophisticated statistical analysis of the full-orbit dynamics as orbits intersect the Poincar\'{e} section for two different values of $\epsilon$.

\subsection{Kruskal Identity}

The purpose of the present paper is to investigate the non-perturbative definition of the guiding-center magnetic moment in the case of doubly-symmetric time-independent magnetic geometries, where the two ignorable spatial coordinates $(q^{1},q^{2})$ are associated with their respective invariant canonical momenta $(P_{1},P_{2})$. In this particular case, the Lagrangian dynamics takes place in a reduced two-dimensional space $(q^{3},dq^{3}/dt)$, where the reduced Lagrangian is derived by the Routh reduction procedure\cite{Brizard:2015}
\begin{equation} 
    L_{r}\left(q^{3},\frac{dq^{3}}{dt}\right) \;\equiv\; L\left(q^{i},\frac{dq^{i}}{dt}\right) \;-\; P_{1}\,\frac{dq^{1}}{dt} \;-\; 
    P_{2}\,\frac{dq^{2}}{dt} \;=\; \frac{1}{2}\,m\,\left(\frac{dq^{3}}{dt}\right)^{2} - V(q^{3};P_{1},P_{2}),
    \label{eq:Routh}
\end{equation}
from which the reduced Euler-Lagrange equations are derived:
\begin{eqnarray*} 
    P_{3} &\equiv& \partial L_{r}/\partial(dq^{3}/dt) \;=\; m\,dq^{3}/dt, \\
    dP_{3}/dt &\equiv& \partial L_{r}/\partial q^{3} \;=\; -\,\partial V/\partial q^{3}. 
\end{eqnarray*}
In addition to the energy-momentum conservation laws for $(E,P_{1},P_{2})$, where the total energy is defined as
\[ E \;\equiv\; \frac{1}{2}\,m\,(dq^{3}/dt)^{2} \;+\; V(q^{3};P_{1},P_{2}), \]
there exists a reduced action integral\cite{Goldstein_2002}
\begin{equation}
    J_{r}(E;P_{1},P_{2}) \;\equiv\; \frac{1}{2\pi}\oint P_{3}\;dq^{3} \;=\; \frac{1}{2\pi}\oint \sqrt{2m\,[E - V(q^{3};P_{1},P_{2})]}\;dq^{3},
    \label{eq:J3_def}
\end{equation}
which is a constant of the reduced motion. In the spirit of Kruskal's pioneering work\cite{Kruskal_1962}, and following recent works by Burby {\it et al.}\cite{Burby_loops_2019,burby_nearly_2023}, the reduced action integral (\ref{eq:J3_def}) is said to satisfy the {\it Kruskal identity}:
\begin{equation} 
    J_{r}(E,P_{1},P_{2}) \;=\; \frac{1}{2\pi}\;\int_{0}^{2\pi} P_{3}\,\pd{q^{3}}{\zeta}\;d\zeta \;\equiv\; \mu_{\rm gc},
    \label{eq:Kruskal}
\end{equation}
where the ordering parameter $\epsilon$ appears in a non-perturbative setting in the definition of $J_{r}$, which yields a non-perturbative definition of the magnetic moment $\mu_{\rm gc}$. We note that since the left-hand side of Eq.~\eqref{eq:Kruskal} is explicitly gyroangle-independent, the guiding-center magnetic moment \eqref{eq:mugc_series}, which is defined in particle phase space, must be gyroangle-independent as well when the full-orbit particle solution is inserted. More importantly, this gyroangle independence must occur without gyroangle-averaging. Hence, a second important constraint associated with the Kruskal identity \eqref{eq:Kruskal} is that the expansion of the guiding-center magnetic moment \eqref{eq:mugc_series} must be explicitly gyroangle-independent, which yields a consistency constraint between the full-orbit particle dynamics and the guiding-center transformation.

In a recent paper \cite{Holas_etal_2025}, the Kruskal identity (\ref{eq:Kruskal}) was proved for two doubly-symmetric magnetic geometries. First, this identity was proved up to fourth order in $\epsilon$ (including second-order effects in magnetic field non-uniformity) for the case of a straight magnetic field ${\bf B} = (1+y)\,\wh{\sf z}$ with a constant gradient\cite{Brizard_2017,Brizard_2022}, where the canonical momenta $(P_{x},P_{z})$ in the ignorable directions $(x,z)$ are invariant. This case is briefly reviewed in App.~\ref{sec:screw_pinch}, where the Kruskal identity \eqref{eq:Kruskal} is proved and the constraint that the gyroangle independence of the guiding-center magnetic moment \eqref{eq:mugc_series} is verified.

Second, a screw-pinch magnetic field was considered \cite{Holas_etal_2025}, with the magnetic field expressed in terms of an Eulerian representation 
\begin{equation} 
{\bf B} \;=\; \nabla\psi\btimes\nabla\theta \;-\; \nabla\Psi(\psi)\btimes\nabla z
\label{eq:B_Euler-Intro}
\end{equation}
in cylindrical geometry, where the axial magnetic flux $\psi(r)$ and the azimuthal magnetic flux $\Psi(\psi)$ are assumed to be functions of the radial position $r$ only, and the canonical momenta $(P_{\theta},P_{z})$ are constants of motion. Unfortunately, the proof of Kruskal's identity (\ref{eq:Kruskal}) for the screw-pinch magnetic field (\ref{eq:B_Euler-Intro}) required considerable use of computer algebra up to third order in $\epsilon$ (i.e., including first-order corrections in magnetic non-uniformity). The main problem associated with an explicit proof of the Kruskal identity \eqref{eq:Kruskal} for this case is that the solution for the azimuthal magnetic flux $\Psi(\psi)$ cannot be obtained in closed form, which prevented the Lagrangian formalism from producing explicit results. The main purpose of the present work is to show that, in contrast to the Lagrangian formalism adopted by Holas {\it et al.}\cite{Holas_etal_2025}, the adoption of the equivalent Newtonian formalism will yield an explicit proof of the Kruskal identity \eqref{eq:Kruskal}, and show that the expansion of the guiding-center magnetic moment \eqref{eq:mugc_series} is explicitly gyroangle-independent.

\subsection{Organization of the paper}

In the present work, we consider the cylindrical screw-pinch magnetic geometry (\ref{eq:B_Euler-Intro}), now expressed as 
\begin{equation}
    {\bf B} \;=\; B(\Theta)\;\left( \cos\Theta\;\wh{\sf z} \;+\; \sin\Theta\;\wh{\theta} \right),
    \label{eq:B_Theta}
\end{equation} 
where the angle $\Theta(r)$ denotes the radial pitch of the screw-pinch magnetic field, and $B(\Theta)$ denotes the magnitude of the magnetic field. Using the Frenet-Serret curvature and torsion coefficients\cite{Brizard:2015} associated with the magnetic unit vector 
\begin{equation}
    \bhat(r,\theta) \;=\; \cos\Theta(r)\;\wh{\sf z} \;+\; \sin\Theta(r)\;\wh{\theta}(\theta),
    \label{eq:b_hat}
\end{equation} 
as well as additional related geometric coefficients that appear in guiding-center perturbation theory\cite{Littlejohn_1983}, we explicitly express the full-orbit Newtonian and Lagrangian mechanics in terms of these geometric coefficients and then construct the reduced radial action integral $J_{r}(E,P_{\theta},P_{z})$, which is defined as an exact invariant of the reduced radial dynamics derived from the reduced Lagrangian $L_{r}(r,dr/dt) = \frac{1}{2}\,(dr/dt)^{2} - V(r;P_{\theta},P_{z})$, where $V(r;P_{\theta},P_{z})$ denotes the reduced radial potential derived by the Routh reduction procedure (\ref{eq:Routh}).

The remainder of the paper is organized as follows. In Sec.~\ref{sec:screw}, we derive the Frenet-Serret curvature and torsion coefficients, as well as other geometric coefficients, associated with the spatial curve (\ref{eq:b_hat}). In Sec.~\ref{sec:Newton}, we will use these geometric coefficients to express Newton's equations of motion in cylindrical screw-pinch magnetic geometry, where we will also introduce particle's pitch angle $\lambda \equiv \cos^{-1}(v_{\|}/v)$ and the particle's gyroangle $\zeta$. In Sec.~\ref{sec:Lagrange}, we apply the Lagrangian formalism to review the particle motion in the doubly-symmetric screw-pinch magnetic geometry from which we derive the reduced radial action $J_{r}(E,P_{\theta},P_{z})$ as an exact particle invariant. 

For the purpose of explicitly proving the Kruskal identity (\ref{eq:Kruskal}) for a doubly-symmetric screw-pinch magnetic geometry, we begin with the Lie-transform perturbation expansion (\ref{eq:mugc_series}) of the guiding-center magnetic moment in Sec.~\ref{sec:mu}. Here, the guiding-center transformation for the particle motion in screw-pinch magnetic geometry, which is reviewed in App.~\ref{sec:gc}, is used to obtain an explicit expression for the guiding-center magnetic moment $\mu_{\rm gc}$ up to first order in magnetic non-uniformity (i.e., third order in the ordering parameter $\epsilon$). In Sec.~\ref{sec:action}, we complete the proof of the Kruskal identity (\ref{eq:Kruskal}) by expanding the reduced radial action $J_{r}(E;P_{\theta},P_{z})$ in powers of $\epsilon$ and perform explicit gyroangle averages to show that $J_{r} \equiv \mu_{\rm gc}$ up to third order in $\epsilon$. Hence, as expected, the guiding-center magnetic moment $\mu_{\rm gc} = J_{r}$ is defined as an exact radial action invariant that depends non-perturbatively on the ordering parameter $\epsilon$. Finally, in Sec.~\ref{sec:conclusion}, we discuss how the guiding-center transformation used in Secs.~\ref{sec:mu}-\ref{sec:action} can be extended to fourth order in $\epsilon$ (i.e., second order in magnetic-field non-uniformity). 

\section{\label{sec:screw}Screw Pinch Magnetic Geometry}

From the screw-pinch magnetic-field unit vector (\ref{eq:b_hat}), we introduce the spatial derivative along a magnetic-field line $d/ds \equiv \bhat\bdot\nabla = r^{-1}\sin\Theta\;\partial/\partial\theta$ (taking into account the $z$-symmetry), and we obtain the Frenet-Serret equations\cite{Brizard:2015} associated with the unit-vector triad $(\bhat,\wh{\sf e}_{1},\wh{\sf e}_{2} \equiv \bhat\btimes\wh{\sf e}_{1})$:
\begin{eqnarray}
    \frac{d\bhat}{ds} &=& \frac{\sin\Theta}{r}\;\pd{\bhat(r,\theta)}{\theta} \;=\; \kappa(r)\;\wh{\sf e}_{1}, \\
    \frac{d\wh{\sf e}_{1}}{ds} &=&  \frac{\sin\Theta}{r}\;\pd{\wh{\sf e}_{1}(r,\theta)}{\theta}  \;=\; -\,\kappa(r)\;\bhat \;+\; \tau(r)\;\wh{\sf e}_{2}, \\
    \frac{d\wh{\sf e}_{2}}{ds} &=& \frac{\sin\Theta}{r}\;\pd{\wh{\sf e}_{2}(r,\theta)}{\theta}  \;=\; -\,\tau(r)\;\wh{\sf e}_{1},
\end{eqnarray}
where the normal (radial) and binormal (transverse) unit vectors are 
\begin{equation}
 \wh{\sf e}_{1} \;\equiv\; -\,\wh{\sf r} \;\;{\rm and}\;\;  \wh{\sf e}_{2} \;=\; -\,\cos\Theta\;\wh{\theta} \;+\; \sin\Theta\;\wh{\sf z}, 
 \label{eq:e_12}
\end{equation}
respectively, and the Frenet-Serret curvature and torsion in screw-pinch magnetic geometry are
\begin{equation}
    \left(\kappa,\; \tau\right) \;\equiv\; \left(\frac{1}{r}\,
    \sin^{2}\,\Theta,\; \frac{1}{r}\,\sin\,\Theta\,\cos\,\Theta\right).
    \label{eq:FS}
\end{equation}
Next, we note that the curl of $\bhat$ is
\begin{eqnarray}
    \nabla\btimes\bhat &=& \nabla\Theta\btimes \left(-\;\sin\Theta\;\wh{\sf z} \;+\; \cos\Theta\;\wh{\theta}\right) \;+\; \sin\Theta\;\nabla\theta\btimes\pd{\wh{\theta}}{\theta} \nonumber \\
     &=& \Theta^{\prime}\;\bhat \;+\; \frac{\wh{\sf z}}{r}\;\sin\Theta \;=\; \tau_{m}\;\bhat \;+\; \kappa\;\wh{\sf e}_{2},
\end{eqnarray}
where the magnetic twist is expressed as 
\begin{equation}
 \tau_{m} \;\equiv\; \bhat\bdot\nabla\btimes\bhat \;=\; \tau \;+\; \Theta^{\prime},
 \label{eq:twist}
\end{equation}
in terms of the Frenet-Serret torsion $\tau(r)$ and the magnetic-pitch shear $\Theta^{\prime}(r)$. 

Lastly, we will also need to evaluate the gyrogauge vector\cite{Littlejohn_1983} in screw-pinch magnetic geometry:
\begin{eqnarray}
    {\bf R} &\equiv& \nabla\wh{\sf e}_{1}\bdot\wh{\sf e}_{2} \;=\; -\,\left(\wh{\theta}\bdot\wh{\sf e}_{2}\right)\,\nabla\theta \;=\; \cos\Theta\;\nabla\theta \;=\; \frac{\cos\Theta}{r}\;\left(\sin\Theta\;\bhat \;-\; \cos\Theta\;\wh{\sf e}_{2}\right) \nonumber \\
    &\equiv& \tau\,\bhat \;-\; \left(\frac{1}{r} - \kappa\right)\;\wh{\sf e}_{2} \;\equiv\; \tau(r)\,\bhat \;-\; R_{\bot}(r)\,\wh{\sf e}_{2},
    \label{eq:gyrogauge}
\end{eqnarray}
which will appear as part of the geometric formulation of Newtonian full-particle dynamics.

\section{\label{sec:Newton}Geometric Newtonian Mechanics}

In this Section, we write Newton's equations of motion $d{\bf r}/dt = {\bf v}$ and $d{\bf v}/dt = (e/m)\,{\bf v}\btimes{\bf B}$ in screw-pinch magnetic geometry for a particle with mass $m$ and charge $e$. Before we do this, however, we want to proceed with dimensionless variables, by normalizing the particle position ${\bf r} \equiv L_{0}\,\ov{\bf r}$, the particle velocity ${\bf v} \equiv v_{0}\,\ov{\bf v}$, and time $t \equiv (m/eB_{0})\,\ov{t}$ (i.e., time is normalized to the gyroperiod). Hence, the dimensionless equations of motion become 
\begin{equation} 
\left. \begin{array}{rcl}
    d\ov{\bf r}/d\ov{t} &=& \epsilon\,\ov{\bf v} \\
    d\ov{\bf v}/d\ov{t} &=& \ov{\bf v}\btimes\ov{\bf B} 
\end{array} \right\},
\label{eq:Newton_epsilon}
\end{equation}
where $\ov{\bf B} \equiv {\bf B}/B_{0}$ is the normalized magnetic field, and $\epsilon \equiv (mv_{0}/eB_{0})/L_{0}$ is the ratio of the characteristic gyroradius $(mv_{0}/eB_{0})$ to the magnetic-nonuniformity length scale $L_{0}$. We note, here, that the characteristic range for $\epsilon$ is $\epsilon \simeq 2\,\times\,10^{-3}$ for a 10 keV proton and $\epsilon \simeq 0.08$ for a 3.5 MeV alpha particle in a 5T magnetic field with $L_{0} = 100$ cm. In what follows, we shall simply assume that $\epsilon \ll 1$ can be used as a suitable perturbation parameter. In addition, we will use the notation $(t,{\bf r},{\bf v})$, without the overbar, to denote the normalized time, particle position, and particle velocity.

\subsection{Velocity representation in screw-pinch magnetic geometry}

Using the unit vectors $(\bhat,\wh{\sf e}_{1},\wh{\sf e}_{2})$ associated with screw-pinch magnetic geometry, the normalized particle velocity can be expressed as
\begin{eqnarray}
    \epsilon\,{\bf v} &=& \frac{dr}{dt}\,\wh{r} \;+\; r\,\frac{d\theta}{dt}\,\wh{\theta} \;+\; \frac{dz}{dt}\,\wh{\sf z} \label{eq:velocity} \\
    &\equiv& \left( r\,\frac{d\theta}{dt}\,\sin\,\Theta \;+\; \frac{dz}{dt}\,\cos\,\Theta\right)\bhat \;-\; \frac{dr}{dt}\,\wh{\sf e}_{1} \;+\;\left( \frac{dz}{dt}\,\sin\,\Theta \;-\; r\,\frac{d\theta}{dt}\,\cos\,\Theta\right) \wh{\sf e}_{2}. \nonumber
\end{eqnarray}
We may also express the particle velocity (\ref{eq:velocity}) as 
${\bf v} \equiv v_{\|}\,\bhat + v_{\bot}\,\wh{\bot}$ in terms of the rotating unit-vector triad $(\bhat,\wh{\bot},\wh{\rho} \equiv \bhat\btimes\wh{\bot})$:\cite{Littlejohn_1983}
\begin{eqnarray}
    \wh{\bot} &\equiv& \pd{\wh{\rho}}{\zeta} \;=\; -\,\sin\zeta\;\wh{\sf e}_{1} \;-\; \cos\zeta\;
    \wh{\sf e}_{2}, \label{eq:bot_def} \\
    \wh{\rho} &\equiv& -\,\pd{\wh{\bot}}{\zeta} \;=\; \cos\zeta\;\wh{\sf e}_{1} \;-\; \sin\zeta\;
    \wh{\sf e}_{2}, \label{eq:rho_def} 
\end{eqnarray}
which are defined in terms of the perpendicular vectors (\ref{eq:e_12}). Here, the parallel velocity $v_{\|}$ is defined as
\begin{equation}
    \epsilon\,v_{\|} \;\equiv\; {\bf v}\bdot\bhat \;=\; r\,\frac{d\theta}{dt}\,\sin\,\Theta \;+\; \frac{dz}{dt}\,\cos\,\Theta,
    \label{eq:v_par}
\end{equation}
while the magnitude of the perpendicular velocity $v_{\bot}$ is defined as
\begin{equation}
    \epsilon\,v_{\bot} \;\equiv\; {\bf v}\bdot\wh{\bot} \;=\; \frac{dr}{dt}\,\sin\zeta \;-\; \left( \frac{dz}{dt}\,\sin\,\Theta \;-\; r\,\frac{d\theta}{dt}\,\cos\,\Theta\right)\,\cos\zeta,
\end{equation}
which can be inverted to yield
\begin{eqnarray}
    \frac{dr}{dt} &\equiv& -\,\epsilon\;{\bf v}\bdot\wh{\sf e}_{1} \;=\; \epsilon\,v_{\bot}\,\sin\zeta, \label{eq:r_dot} \\
    \frac{dz}{dt}\,\sin\,\Theta \;-\; r\,\frac{d\theta}{dt}\,\cos\,\Theta &\equiv& \epsilon\;{\bf v}\bdot\wh{\sf e}_{2} \;=\; -\;\epsilon\,v_{\bot}\,\cos\zeta, \label{eq:v_perp}
\end{eqnarray}
since $v_{\bot}$ must be independent of the gyroangle $\zeta$. Using Eqs.~(\ref{eq:v_par}) and (\ref{eq:v_perp}), the cylindrical velocity components $(d\theta/dt, dz/dt)$ are, therefore, expressed in screw-pinch magnetic geometry as
\begin{eqnarray}
    r\;\frac{d\theta}{dt} &=& \epsilon\,v_{\|}\;\sin\,\Theta \;+\; \epsilon\,v_{\bot}\;\cos\,\Theta\,\cos\zeta, \label{eq:theta_dot} \\
    \frac{dz}{dt} &=& \epsilon\,v_{\|}\;\cos\,\Theta \;-\; \epsilon\,v_{\bot}\;\sin\,\Theta\,\cos\zeta. \label{eq:z_dot}
\end{eqnarray}
Here, the cylindrical velocity components (\ref{eq:r_dot}) and (\ref{eq:theta_dot})-(\ref{eq:z_dot}) are expressed in screw-pinch magnetic geometry in terms of the pitch angle $\Theta$ and the velocity coordinates $(v_{\|}, v_{\bot},\zeta)$.

\subsection{Newton's equations of motion in screw-pinch magnetic geometry}

Newton's equation of motion for the normalized particle velocity (\ref{eq:velocity}) is
\begin{equation}
    \frac{d{\bf v}}{dt} \;=\; \frac{dv_{\|}}{dt}\,\bhat \;+\; v_{\|}\;
    \frac{d\bhat}{dt} \;+\; \frac{dv_{\bot}}{dt}\,\wh{\bot} \;+\; v_{\bot}\;\frac{d\wh{\bot}}{dt} \;=\; B\,{\bf v}\btimes\bhat \;=\; -\;v_{\bot}\,B\;\wh{\rho},
    \label{eq:Newton}
\end{equation}
where
\begin{eqnarray}
    \frac{d\bhat}{dt} &=& \frac{d\theta}{dt}\,\sin\,\Theta\;\wh{\sf e}_{1} \;-\; \frac{d\Theta}{dt}\,\wh{\sf e}_{2}, 
    \nonumber \\
     &=& \epsilon\,\left( v_{\|}\,\kappa \;+\; v_{\bot}\,\tau\,\cos\zeta\right)\,\wh{\sf e}_{1} \;-\; \epsilon\,\left(\Theta^{\prime}\;v_{\bot}\,\sin\zeta\right)\;\wh{\sf e}_{2}, \label{eq:d_bhat} \\
    \frac{d\wh{\bot}}{dt} &=& \left( \frac{d\theta}{dt}\,\sin\,\Theta\,\sin\zeta \;-\; \frac{d\Theta}{dt}\,\cos\zeta \right) \bhat \;-\;
    \left( \frac{d\zeta}{dt} \;-\; \frac{d\theta}{dt}\,\cos\,\Theta\right) \wh{\rho} \nonumber \\
    &=& \epsilon\,\sin\zeta\;\left[ v_{\|}\,\kappa \;+\; v_{\bot}\,\left(\tau - \Theta^{\prime}\right)\,\cos\zeta\right]\bhat \;-\; \left[ \frac{d\zeta}{dt} \;-\; 
    \epsilon\,\left( v_{\|}\,\tau \;+\; v_{\bot}\,R_{\bot}\;\cos\zeta\right)\right] \wh{\rho}.
    \label{eq:d_bot}
\end{eqnarray}
In deriving Eq.~(\ref{eq:d_bot}), we used
\begin{eqnarray}
    \frac{d\wh{\sf e}_{1}}{dt} &=& -\,\frac{d\theta}{dt}\;\wh{\theta} \;=\; -\,\frac{d\theta}{dt}\;\left(\sin\,\Theta\,\bhat \;-\; \cos\,\Theta\,
    \wh{\sf e}_{2}\right), \\
    \frac{d\wh{\sf e}_{2}}{dt} &=& \frac{d\Theta}{dt}\;\bhat \;-\; \frac{d\theta}{dt}\,\cos\,\Theta\;\wh{\sf e}_{1}.
\end{eqnarray}
Equation (\ref{eq:Newton}), therefore, yields the three coupled equations for the velocity coordinates:
\begin{eqnarray}
    \frac{dv_{\|}}{dt} &=& -\,v_{\bot}\,\frac{d\wh{\bot}}{dt}\bdot\bhat \;=\; -\,\epsilon\,v_{\bot}\,\sin\zeta \left[ v_{\|}\,\kappa \;+\; v_{\bot}\,\left(\tau - \Theta^{\prime}\right)\,\cos\zeta\right], \label{eq:vpar_dot} \\
    \frac{dv_{\bot}}{dt} &=& -\,v_{\|}\,\frac{d\bhat}{dt}\bdot\wh{\bot} \;=\; \epsilon\,v_{\|}\,\sin\zeta \left[ v_{\|}\,\kappa \;+\; v_{\bot}\,\left(\tau - \Theta^{\prime}\right)\,\cos\zeta\right], \label{eq:vperp_dot} \\
    \frac{d\zeta}{dt} &=& B \;+\; \epsilon\,v_{\|}\,\left[ \tau_{m} + \left(\tau - \Theta^{\prime}\right)\,\cos^{2}\zeta\right] \;+\; \epsilon\,v_{\bot}\,\cos\zeta \left( R_{\bot} \;+\; \frac{v_{\|}^{2}}{v_{\bot}^{2}}\;\kappa\right), \label{eq:zeta_dot}
\end{eqnarray}
where the Frenet-Serret coefficients $(\kappa,\tau,\tau_{m})$ are given by Eqs.~(\ref{eq:FS}) and (\ref{eq:twist}) and the gyrogauge component $R_{\bot} \equiv 1/r - \kappa$ is given by Eq.~(\ref{eq:gyrogauge}). Using Eqs.~(\ref{eq:r_dot}) and (\ref{eq:vperp_dot}), we note that, while the time derivative of the lowest-order magnetic moment $\mu_{0} \equiv \frac{1}{2} \epsilon^{2}v_{\bot}^{2}/B(\Theta)$:
\begin{eqnarray} 
    \frac{d\mu_{0}}{dt} &=& \frac{\epsilon^{2}v_{\bot}}{B(\Theta)}\;\frac{dv_{\bot}}{dt} \;-\; \frac{\epsilon^{2}\,v_{\bot}^{2}\,B^{\prime}(\Theta)}{2\,B^{2}(\Theta)}\;\frac{d\Theta}{dt} \nonumber \\
    &=& \frac{\epsilon^{3}\,v_{\|}\,v_{\bot}}{B(\Theta)}\;\sin\zeta \left[ v_{\|}\,\kappa \;+\; v_{\bot}\,(\tau - \Theta^{\prime})\,\cos\zeta\right] \;-\;
    \frac{\epsilon^{3}\,v_{\bot}^{3}\,B^{\prime}(\Theta)}{2\,B^{2}(\Theta)}\;\Theta^{\prime}\;\sin\zeta,
\end{eqnarray}
does not vanish, its gyroangle average $\langle d\mu_{0}/dt\rangle = 0$ vanishes at the lowest order, which is the hallmark of adiabatic invariance\cite{Northrop_1963}.

Lastly, we note that the kinetic energy $E = \epsilon^{2}(v_{\|}^{2} + v_{\bot}^{2})/2$ is a constant of motion, since $dE/dt = \epsilon^{2}\,v_{\|}\,dv_{\|}/dt + \epsilon^{2}\,v_{\bot }\,dv_{\bot}/dt = 0$ when Eqs.~\eqref{eq:vpar_dot}-\eqref{eq:vperp_dot} are taken into account. The parallel velocity $v_{\|} = v\,\cos\lambda$ and the perpendicular velocity $v_{\bot} = v\,\sin\lambda$ can, therefore, be expressed in terms of the pitch angle $\lambda$, so that the equation of motion for the pitch angle $\lambda$ is expressed as
\begin{equation}
    \frac{d\lambda}{dt} \;=\; \frac{1}{v_{\|}}\,\frac{dv_{\bot}}{dt} \;=\; \epsilon\,\sin\zeta \left[ v_{\|}\,\kappa \;+\; v_{\bot}\,\left(\tau - \Theta^{\prime}\right)\,\cos\zeta\right].
    \label{eq:lambda_dot}
\end{equation}
Hence, the equations of motion for the cylindrical coordinates $(r,\theta,z)$ and the velocity-space angles $(\lambda,\zeta)$ are expressed in terms of the screw-pinch magnetic geometric coefficients $(\kappa,\tau,\tau_{m},R_{\bot})$.

\section{\label{sec:Lagrange}Particle Lagrangian Mechanics}

As an alternative to the geometric Newtonian representation presented in Sec.~\ref{sec:Newton}, we now explore the Lagrangian representation of the particle dynamics in the screw-pinch magnetic field (\ref{eq:B_Theta}). We now choose the magnitude of the magnetic field to be defined as
\begin{equation}
    B(\Theta) \;\equiv\; \sec\Theta.
    \label{eq:B_mag}
\end{equation}
We note that, while the choice (\ref{eq:B_mag}) allows us to obtain explicit  expressions, it does not impact on the main results of our work. While the Lagrangian formalism yields identical results obtained from the Newtonian formalism adopted in Sec.~\ref{sec:Newton}, it is presented, here, as a way to derive the reduced radial action integral [see Eq.~\eqref{eq:Jr_def}], which can then be inserted in the Kruskal identity \eqref{eq:Kruskal} for the screw-pinch magnetic geometry.

\subsection{Euler-potential magnetic representation}

In order to proceed with a Lagrangian representation, using the magnitude (\ref{eq:B_mag}), the magnetic field (\ref{eq:B_Theta}) can also be expressed in Euler form as\cite{Holas_etal_2025}
\begin{equation}
    {\bf B} \;\equiv\; B(\Theta)\;\bhat \,=\, \wh{\sf z} + \tan\Theta\;\wh{\theta} \;=\; \nabla\psi(r)\btimes\nabla\theta \;-\; \nabla\Psi(r)\btimes\nabla z, \label{eq:B_Euler}
\end{equation}
where the dimensonless magnetic fluxes $\psi(r)$ and $\Psi(r)$ satisfy the differential equations
\begin{equation}
    \frac{d\psi(r)}{dr} \;=\; r \;\;{\rm and}\;\; \frac{d\Psi(r)}{dr} \;=\; \tan\Theta.
    \label{eq:psi_Psi}
\end{equation}
The solution for $\psi(r)$ is easily solved as $\psi(r) = r^{2}/2$, which is chosen to vanish at $r = 0$. Lastly, we note that it is a common practice to write $\Psi(\psi)$, which leads to the differential equation for the rotational transform\cite{Holas_etal_2025} $\Psi^{\prime}(\psi) = r^{-1}\;\tan\Theta$.
In what follows, however, we will not need to solve for $\Psi(\psi)$.

\subsection{Particle Lagrangian}

Using the normalized magnetic vector potential ${\bf A} = \psi\;\nabla\theta - \Psi\;\nabla z$ associated with the magnetic field (\ref{eq:B_Euler}), we express the particle Lagrangian in cylindrical coordinates as
\begin{eqnarray}
    L &=& \left({\bf A} \;+\; \frac{d{\bf r}}{dt}\right) \bdot\frac{d{\bf r}}{dt} \;-\; \frac{1}{2}\,\left|\frac{d{\bf r}}{dt}\right|^{2} \nonumber \\
    &=& \left(\psi(r)\,\frac{d\theta}{dt} \;-\; \Psi(r)\,\frac{dz}{dt}\right) \;+\; \frac{1}{2} \left[ \left(\frac{dr}{dt}\right)^{2} + r^{2}\,\left(\frac{d\theta}{dt}\right)^{2} + \left(\frac{dz}{dt}\right)^{2} \right].
    \label{eq:lag_def}
\end{eqnarray}

From this Lagrangian, we obtain the canonical momenta
\begin{eqnarray}
    P_{r} &\equiv& \pd{L}{(dr/dt)} \;=\; \frac{dr}{dt}, \\
    P_{\theta} &\equiv& \pd{L}{(d\theta/dt)} \;=\; \psi \;+\; r^{2}\,\frac{d\theta}{dt}, 
    \label{eq:P_theta} \\
    P_{z} &\equiv& \pd{L}{(dz/dt)} \;=\; -\,\Psi \;+\; \frac{dz}{dt},
    \label{eq:P_z}
\end{eqnarray}
and the radial Euler-Lagrange equation yields
\begin{eqnarray}
    \frac{dP_{r}}{dt} \;=\; \frac{d^{2}r}{dt^{2}} \;=\; \pd{L}{r} &=& \frac{d\psi(r)}{dr}\;\frac{d\theta}{dt} \;-\; \frac{d\Psi(r)}{dr}\;\frac{dz}{dt} 
    \;+\; r\;\left(\frac{d\theta}{dt}\right)^{2} \nonumber \\
     &=&r\;\frac{d\theta}{dt} \;-\; \tan\Theta\;\frac{dz}{dt} \;+\; r\;\left(\frac{d\theta}{dt}\right)^{2},
    \label{eq:dP_r}
\end{eqnarray}
where we used Eq.~(\ref{eq:psi_Psi}). The remaining Euler-Lagrange equations are $dP_{\theta}/dt = \partial L/\partial\theta \equiv 0$, and $dP_{z}/dt = \partial L/\partial z \equiv 0$, i.e., because of the symmetries of the Lagrangian (\ref{eq:lag_def}) with respect to $(\theta,z)$, the canonical momenta $(P_{\theta},P_{z})$ are constants of motion. 

If we substitute Eqs.~(\ref{eq:theta_dot})-(\ref{eq:z_dot}) into Eq.~(\ref{eq:dP_r}), we obtain the radial equation
\begin{eqnarray}
    \frac{d^{2}r}{dt^{2}} &=& \epsilon\,v_{\|}\;\sin\,\Theta \;+\; \epsilon\,v_{\bot}\;\cos\,\Theta\,\cos\zeta \;-\; \epsilon\,v_{\|}\;\sin\,\Theta \;+\; \epsilon\,v_{\bot}\;\cos\Theta\,(\sec^{2}\Theta - 1)\,\cos\zeta \nonumber \\
    &&+\; \frac{\epsilon^{2}}{r} \left(v_{\|}\;\sin\,\Theta \;+\; v_{\bot}\;\cos\,\Theta\,\cos\zeta\right)^{2} \nonumber \\
    &=& \epsilon\,v_{\bot}\;B\;\cos\zeta \;+\; \epsilon^{2} \left( v_{\|}^{2}\;\kappa \;+\; v_{\bot}^{2}\,R_{\bot}\,\cos^{2}\zeta \;+\; 2\;
    v_{\|}\,v_{\bot}\,\tau\;\cos\zeta\right).
\end{eqnarray}
Next, if we insert Eq.~(\ref{eq:zeta_dot}) for $d\zeta/dt$, we obtain
\begin{eqnarray}
    \frac{d^{2}r}{dt^{2}} &=& \epsilon\,v_{\bot}\;\cos\zeta\;\frac{d\zeta}{dt} \;+\; \epsilon^{2}\,v_{\|}\,\sin^{2}\zeta \left[ v_{\|}\;\kappa \;+\; v_{\bot}\,\left(\tau - \Theta^{\prime}\right)\,\cos\zeta\right] \nonumber \\
    &=& \frac{d}{dt}\left(\epsilon\,v_{\bot}\,\sin\zeta\right) \;\equiv\; \frac{d^{2}r}{dt^{2}},
\end{eqnarray}
which follows after substituting Eq.~(\ref{eq:vperp_dot}) for $dv_{\bot}/dt$. Hence, as expected, the Euler-Lagrange equations exactly agree with Newton's equations. While this result appears trivial, it serves as an explicit proof of the validity of the geometric Newtonian formalism introduced in Sec.~\ref{sec:Newton}.

\subsection{Reduced radial particle dynamics}

In what follows, we will use the notation $P_{\theta} \equiv \ov{\psi}$ and $P_{z} = -\,\ov{\Psi}$ for the canonical-momentum constants of motion, so that Eqs.~(\ref{eq:P_theta})-(\ref{eq:P_z}) yield
\begin{eqnarray}
    \psi(r) &=& \ov{\psi} \;-\; r^{2}\frac{d\theta}{dt}, \label{eq:psi_ovpsi} \\
    \Psi(r) &=& \ov{\Psi} \;+\; \frac{dz}{dt}. \label{eq:Psi_ovPsi}
\end{eqnarray}
Because the Lagrangian (\ref{eq:lag_def}) is independent of the coordinates $(\theta,z)$, we may use the Routh reduction procedure\cite{Brizard:2015} to obtain the reduced radial Lagrangian
\begin{equation}
    L_{r}(r,dr/dt) \;\equiv\; L \;-\; \frac{d\theta}{dt}\,P_{\theta} \;-\; \frac{dz}{dt}\,P_{z} \;=\; \frac{1}{2}\,\left(\frac{dr}{dt}\right)^{2} \;-\; V(r),
\end{equation}
where the reduced radial potential is
\begin{equation}
    V(r) \;=\; \frac{1}{2}\,\left[ \left(\Psi(\psi) - \ov{\Psi}\right)^{2} \;+\; \frac{1}{r^{2}}\,\left(\psi(r) - \ov{\psi}\right)^{2}\right].
    \label{eq:Vr_screw}
\end{equation}
The reduced radial equation of motion is $d^{2}r/dt^{2} = -\,V^{\prime}(r)$, and the law of energy conservation is
\begin{equation}
    E \;=\; \frac{1}{2}\,\left(\frac{dr}{dt}\right)^{2} \;+\; V(r),
\end{equation}
The reduced radial action integral\cite{Goldstein_2002}
\begin{equation}
    J_{r}(E,P_{\theta},P_{z}) \;\equiv\; \frac{1}{2\pi} \oint P_{r}\,dr
    \;=\; \frac{1}{2\pi}\int_{0}^{2\pi} \sqrt{2\left[ E \;-\; V(r)\right]}\;\frac{dr}{d\zeta}\;d\zeta
    \label{eq:Jr_def}
\end{equation}
is also an invariant of the reduced radial motion. With three separate exact constants of motion $(E,P_{\theta},P_{z})$, the motion of a charged particle in this screw-pinch magnetic field is said to be integrable. In Eq.~\eqref{eq:Jr_def}, using the equivalence of the Newtonian and Lagrangian formalisms, we will express $dr/d\zeta \equiv (dr/dt)/(d\zeta/dt)$ in terms of $dr/dt = \sqrt{2\,[E - V(r)]}$ and $d\zeta/dt$, which are defined respectively by the Newtonian expressions \eqref{eq:r_dot} and \eqref{eq:zeta_dot} . The benefit of the Lagrangian formalism is that it provides an explicit derivation of the reduced radial potential \eqref{eq:Vr_screw}, which yields the reduced radial action integral \eqref{eq:Jr_def}. However, because the azimuthal magnetic flux $\Psi(\psi)$ cannot be solved in closed form, we will show that the geometric Newtonian formalism will yield explicit expressions for the reduced radial action integral \eqref{eq:Jr_def}. We will return to the reduced radial action integral (\ref{eq:Jr_def}) in Sec.~\ref{sec:action} to show that it matches exactly with the magnetic moment up to third order in $\epsilon$.

\section{\label{sec:mu}Guiding-center Magnetic Moment}

In this Section, we will proceed with a calculation of the magnetic moment (expressed in units of action) $\mu_{\rm gc} = \mu_{0} + \epsilon\,\mu_{1} + \cdots$ up to first order in magnetic-field non-uniformity. In the next Section, we will show that $\mu_{\rm gc} \equiv J_{r}$ matches exactly with the asymptotic expansion of the reduced radial action (\ref{eq:Jr_def}).

\subsection{Zeroth-order magnetic moment}

The lowest-order (normalized) magnetic moment $\mu_{0} = |d{\bf r}_{\bot}/dt|^{2}/2B$ is defined as
\begin{equation}
    \mu_{0} \;\equiv\; \frac{\epsilon^{2}v_{\bot}^{2}}{2\,B} \;=\; \frac{\epsilon^{2}}{2}\,v^{2}\;\sin^{2}\lambda\;\cos\Theta,
    \label{eq:mu_0}
\end{equation}
where both $(r,\lambda)$ need to be expanded in powers of $\epsilon$ through the guiding-center transformation: $r \equiv r_{0} + \epsilon\,r_{1} + \cdots$ and $\lambda \equiv \lambda_{0} + \epsilon\,\lambda_{1} + \cdots$, where $(r_{1},\lambda_{1})$ denote first-order corrections, while the speed $v$ is a constant of motion. Hence, up to third order in $\epsilon^{3}$, we find
\begin{eqnarray}
    \mu_{0} &=& \frac{\epsilon^{2}}{2}\;v_{\bot0}^{2}\;\cos\Theta_{0} \;+\; \frac{\epsilon^{3}}{2}\,v_{\bot0} \left[ v_{\|0}\;\left( 2\,\cos\Theta_{0}\right)\;\lambda_{1} \;-\; v_{\bot0}\;\left( \Theta_{0}^{\prime}\,\sin\Theta_{0}\right)\;r_{1}\right] \nonumber \\
    &=& \ov{\mu_{0}}\left( 1 \;-\; \epsilon\;\Theta_{0}^{\prime}\,\tan\Theta_{0}\;r_{1}\right) \;+\;
    \epsilon^{3}\;v_{\|0}\,v_{\bot0}\;\cos\Theta_{0}\;\lambda_{1},
    \label{eq:mu0_final}
\end{eqnarray}
where the lowest-order magnetic moment $\ov{\mu_{0}} \equiv \frac{1}{2}\,\epsilon^{2}\,v_{\bot0}^{2}\cos\Theta_{0}$ is evaluated at the constants of motion $(r_{0},\lambda_{0})$.

\subsection{First-order magnetic moment}

In standard guiding-center theory \cite{Littlejohn_1983,Cary_2009,tronko_lagrangian_2015}, the first-order magnetic moment is
\begin{equation}
    \mu_{1} \;=\; \left(\ov{\mu_{0}}\,\nabla\ln B + \epsilon^{2}\,v_{\|0}^{2}\,\cos\Theta\;\vb{\kappa}\right)\bdot\vb{\rho}_{0} \;-\; 
    \epsilon\,\ov{\mu_{0}}\,v_{\|0}\,\cos\Theta\;(\tau_{m} + \alpha_{1}),
    \label{eq:mu_1def}
\end{equation}
where the lowest-order gyroradius vector is
\begin{eqnarray}
    \vb{\rho}_{0} &=& \frac{\bhat}{B}\btimes\,\epsilon\;{\bf v}_{\bot} \;=\; \epsilon\,v_{\bot0}\,\cos\Theta_{0} \left( \cos\zeta\,
    \wh{\sf e}_{1} \;-\; \sin\zeta\,\wh{\sf e}_{2} \right),
    \label{eq:rho_0}
\end{eqnarray}
and the gyroangle-dependent correction\cite{Littlejohn_1983} $\alpha_{1} \equiv -\frac{1}{2}\,(\wh{\bot}\bdot\nabla\bhat\bdot\wh{\rho} + \wh{\rho}\bdot\nabla\bhat\bdot\wh{\bot})$ can be expressed in screw-pinch magnetic geometry as
\[ \alpha_{1} \;=\; \frac{1}{2}\,(\Theta^{\prime} - \tau)\,\cos(2\zeta) \;=\; \frac{1}{2}\,\tau_{m} \;-\; \left(\tau\;\cos^{2}\zeta \;+\; \Theta^{\prime}\;\sin^{2}\zeta\right), \]
so that the torsional correction
\begin{equation}
    \tau_{m} + \alpha_{1} \;=\; \tau_{0} \left(\frac{3}{2} - \cos^{2}\zeta\right) \;+\; \Theta^{\prime}_{0} \left(\frac{3}{2} - \sin^{2}\zeta\right)
    \label{eq:tau_alpha}
\end{equation}
is expressed in terms of the Frenet-Serret torsion $\tau$ and the magnetic-pitch shear $\Theta^{\prime}$, which are both evaluated at $r_{0}$. Equation (\ref{eq:mu_1def}) may, therefore, be rewritten as
\begin{equation}
    \mu_{1} \;=\; \ov{\mu_{0}}\,\vb{\rho}_{0}\bdot\nabla\ln B \;+\; \epsilon\,\rho_{\|0} \left[ \epsilon^{2}\,v_{\|0}\,\vb{\rho}_{0}\bdot\vb{\kappa} \;-\;
    \ov{\mu_{0}}\;(\tau_{m} + \alpha_{1})\right],
    \label{eq:mu_1}
\end{equation}
where the first term is $\ov{\mu_{0}}\,\vb{\rho}_{0}\bdot\nabla\ln B = -\;\epsilon\,\ov{\mu_{0}}\,\rho_{\bot0}\cos\zeta\;\left(\Theta^{\prime}_{0}\;\tan\Theta_{0}\right)$.

We now use the inverse guiding-center transformation $z^{\alpha} \equiv Z_{\rm gc}^{\alpha} - \epsilon\,G_{1}^{\alpha} + \cdots$ presented in App.~\ref{sec:gc}, which yields
\begin{eqnarray}
    r &=& r_{\rm gc} \;-\; \epsilon\;\wh{\sf r}\bdot G_{1}^{\bf x} \;=\; r_{\rm gc} \;-\; \epsilon\,\wh{\sf e}_{1}\bdot\vb{\rho}_{0} \;\equiv\; r_{0} + \epsilon\,r_{1}, \\
    \lambda &=& \lambda_{\rm gc} \;-\; \epsilon\,G_{1}^{\lambda} \;\equiv\; \lambda_{0} + \epsilon\,\lambda_{1},
\end{eqnarray}
where the first-order corrections are
\begin{eqnarray}
    r_{1} &=& -\;\rho_{\bot}\;\cos\zeta \;=\; -\,v_{\bot0}\,\cos\Theta_{0}\;\cos\zeta, \label{eq:r1_gc} \\
    \lambda_{1} &=& -\;G_{1}^{v_{\|}}/v_{\bot0} \;=\; \left[ \ov{\mu_{0}}\;(\tau_{m} + \alpha_{1}) \;-\; \epsilon\,v_{\|0}\,\vb{\rho}_{0}\bdot\vb{\kappa} 
    \right]/v_{\bot0}, \label{eq:lambda1_gc}
\end{eqnarray}
so that the first-order correction (\ref{eq:mu_1}) becomes
\begin{equation}
    \mu_{1} \;=\; \ov{\mu_{0}}\;\Theta_{0}^{\prime}\,\tan\Theta_{0}\;r_{1} \;-\; \epsilon^{2}\;v_{\|0}\,v_{\bot0}\;\cos\Theta_{0}\;\lambda_{1}.
    \label{eq:mu1_final}
\end{equation}
Hence, by combining Eqs.~\eqref{eq:mu0_final} and \eqref{eq:mu1_final}, we obtain the full magnetic moment (up to third order in $\epsilon$)
\begin{equation}
    \mu_{\rm gc} \;=\; \mu_{0} \;+\; \epsilon\,\mu_{1} \;=\; \ov{\mu_{0}} \;\equiv\; \frac{1}{2}\,\epsilon^{2}\,v_{\bot0}^{2}\,\cos\Theta_{0},
    \label{eq:mu_final}
\end{equation}
which implies that the first-order corrections of $\mu_{0}$ are exactly canceled out by $\mu_{1}$. We note that this cancellation is not unique to the case of the screw-pinch magnetic geometry, as it is also observed in the case of a slab magnetic field\cite{Holas_etal_2025} (see also App.~\ref{sec:screw_pinch} in this paper).

\section{\label{sec:action}Expansion of the Reduced Radial Action}

We now turn our attention to the reduced radial action integral (\ref{eq:Jr_def}), which can be expressed as
\begin{equation}
    J_{r} \;=\; \frac{1}{2\pi}\oint P_{r}\,dr \;=\; \frac{1}{2\pi}\int_{0}^{2\pi} \left(\frac{dr}{dt}\right)^{2}\;\frac{d\zeta}{(d\zeta/dt)},
    \label{eq:Jr_rzeta}
\end{equation}
where the Newtonian equations for $(dr/dt,d\zeta/dt)$ are given by Eqs.~(\ref{eq:r_dot}) and (\ref{eq:zeta_dot}), respectively. In order to compare the reduced radial action integral (\ref{eq:Jr_rzeta}) with the magnetic moment (\ref{eq:mu_final}), we need to expand $(dr/dt)^{2}$ up to third order in $\epsilon$ and $(d\zeta/dt)^{-1}$ up to first order in $\epsilon$. First, we find
\begin{eqnarray}
    (dr/dt)^{2} &=& \epsilon^{2}\,\sin^{2}\zeta \left( v_{\bot0}^{2} \;+\; 2\;\epsilon\,v_{\|0}\,v_{\bot0}\;\lambda_{1}\right), \\
    (d\zeta/dt)^{-1} &=& \cos\Theta_{0} \;-\; \epsilon\;\left(\Theta_{0}^{\prime}\,\sin\Theta_{0}\right)\;r_{1} \;-\; \epsilon\,v_{\|0}\,\cos\Theta_{0} \left[ \tau_{0}\;\left(1 + \cos^{2}\zeta\right) \;+\; \Theta_{0}^{\prime}\;\sin^{2}\zeta \right] \nonumber 
     \\
     &&-\; \epsilon\,v_{\bot0}\,\cos\Theta_{0}\,\cos\zeta \left( R_{\bot0} \;+\; \frac{v_{\|0}^{2}}{v_{\bot0}^{2}}\;\kappa_{0}\right),
\end{eqnarray}
where Eq.~(\ref{eq:lambda1_gc}) yields
\begin{equation}
    \lambda_{1} \;=\; \frac{v_{\bot0}}{2}\,\cos\Theta_{0}\;\left[\tau_{0}\,\left(\frac{3}{2} - \cos^{2}\zeta\right) + \Theta_{0}^{\prime} \left(\frac{3}{2} - \sin^{2}\zeta\right) \right] \;-\; v_{\|0}\,\cos\Theta_{0}\;\kappa_{0}\,\cos\zeta,
\end{equation}
after inserting Eq.~(\ref{eq:tau_alpha}).

Next, when we expand Eq.~(\ref{eq:Jr_rzeta}) up to third order in $\epsilon$, we obtain a collection of four gyroangle-averaged terms
\begin{eqnarray}
    J_{r} &=& \epsilon^{2}\,v_{\bot0}^{2}\,\cos\Theta_{0}\;\left\langle\sin^{2}\zeta\right\rangle \;+\; \epsilon^{3}\,v_{\bot0}^{2}v_{\|0} \left\langle \sin^{2}\zeta \left[ \tau_{0}\,\left(3/2 - \cos^{2}\zeta\right) + \Theta_{0}^{\prime} \left(3/2 - \sin^{2}\zeta\right)\right]\right\rangle 
    \nonumber \\
    &&-\; \epsilon^{3}\,v_{\bot0}^{2}v_{\|0} \left\langle \sin^{2}\zeta \left[ \tau_{0}\,\left(1 + \cos^{2}\zeta\right) + \Theta_{0}^{\prime}\;\sin^{2}\zeta\right]\right\rangle \;-\;
    \epsilon^{3} (\cdots)\;\left\langle \sin^{2}\zeta\,\cos\zeta\right\rangle,
    \label{eq:Jr_gyro}
\end{eqnarray}
where gyroangle averaging is denoted as $\langle\cdots\rangle$ and the last term $(\cdots)$ in Eq.~\eqref{eq:Jr_gyro} combines contributions that are proportional to $\sin^{2}\zeta\,\cos\zeta$. 

Finally, using the identities $\langle\sin^{2}\zeta\rangle = 1/2$ and $\langle\sin^{2}\zeta\,\cos\zeta\rangle = 0$, and the expressions
\begin{eqnarray*}
   \left\langle \sin^{2}\zeta \left[ \tau_{0}\,\left(3/2 - \cos^{2}\zeta\right) + \Theta_{0}^{\prime} \left(3/2 - \sin^{2}\zeta\right)\right]\right\rangle &=& \frac{1}{8}\;\left(5\,\tau_{0} \;+\; 3\,\Theta_{0}^{\prime}\right), \\
   \left\langle \sin^{2}\zeta \left[ \tau_{0}\,\left(1 + \cos^{2}\zeta\right) + \Theta_{0}^{\prime}\;\sin^{2}\zeta\right]\right\rangle &=& \frac{1}{8}\;\left(5\,\tau_{0} \;+\; 3\,\Theta_{0}^{\prime}\right),
\end{eqnarray*}
we obtain the final result from  Eq.~\eqref{eq:Jr_gyro}:
\begin{equation}
    J_{r} \;=\; \frac{1}{2}\,\epsilon^{2}\,v_{\bot0}^{2}\,\cos\Theta_{0} \;=\; \ov{\mu}_{0} \;\equiv\; \mu_{\rm gc},
    \label{eq:Jr_mu}
\end{equation}
which confirms that, up to third order in $\epsilon$, the reduced radial action integral is identical to the magnetic moment \eqref{eq:mu_final}.

\section{\label{sec:conclusion}Conclusion and Discussion}

The present work was concerned with an explicit proof of the Kruskal identity \eqref{eq:Kruskal} in a doubly-symmetric screw-pinch magnetic geometry. In recent work by Holas {\it et al.}\cite{Holas_etal_2025}, a Lagrangian formulation was used to derive an explicit expression for the reduced radial action integral (\ref{eq:Jr_rzeta}), which is expressed in terms of the axial magnetic flux $\psi(r)$ and the azimuthal magnetic flux $\Psi(\psi)$ defined in Eq.~\eqref{eq:psi_Psi}. Unfortunately, since the azimuthal magnetic flux 
$\Psi(\psi)$ cannot be solved in closed form, the Lagrangian formalism adopted by Holas {\it et al.}\cite{Holas_etal_2025} yielded results that could only prove the Kruskal identity through computer algebra up to third order in $\epsilon$.  By using the geometric formulation of the Newtonian mechanics of charged particle motion in a screw-pinch magnetic field in the present work, we have demonstrated Kruskal's identity between the reduced radial action integral (\ref{eq:Jr_rzeta}), from which the Kruskal adiabatic-invariant series is derived, and the magnetic moment (\ref{eq:mu_final}), which is expanded explicitly in powers of the magnetic-field non-uniformity. By exploiting the algorithmic nature of the Lie-transform approach to guiding-center dynamics, which would generate explicit expressions for $\mu_{2}$ and $(r_{2},\lambda_{2})$, it is possible to prove the Kruskal identity $J_{r} = \mu_{\rm gc}$ up to fourth order in $\epsilon$ (i.e., second order in magnetic-field non-uniformity). 

It has recently been argued\cite{Holas_etal_2025} that a non-perturbative expression for the gyrocenter magnetic moment, which would be an adiabatic invariant in the presence of time-dependent electromagnetic-field perturbations, might be derived by the same method used here. While the particle energy is no longer conserved because of the time dependence of the field perturbations, the double spatial symmetries of the unperturbed magnetic field configuration will also be destroyed. Hence, in the absence of a reduced action integral (\ref{eq:J3_def}), one would only rely on the formal integral definition for the gyrocenter magnetic moment
\begin{equation}
    \mu_{\rm gy} \;\equiv\; \frac{\epsilon}{2\pi}\int_{0}^{2\pi}{\bf P}_{\rm gy}\bdot\pd{\vb{\rho}_{\rm gy}}{\zeta}\;d\zeta \;=\; \mu_{\rm gc} \;+\; \delta\,\mu_{\rm gy1} \;+\; \cdots,
    \label{eq:mugy_integral}
\end{equation}
where $\mu_{\rm gc}$ denotes the possibly non-perturbative (unperturbed) guiding-center magnetic moment, the ordering parameter $\delta$ is defined in terms of the amplitude of the field perturbation\cite{Brizard_2007},  and an additional time-dependent transformation
\begin{equation}
    {\sf T}_{\rm gy}^{-1}f \;\equiv\; \cdots {\sf T}_{3}^{-1}{\sf T}_{2}^{-1}{\sf T}_{1}^{-1}f \;=\; f \;-\; \delta\,{\cal L}_{1}f \;-\; \delta^{2} \left( {\cal L}_{2}f \;-\; \frac{1}{2}\,{\cal L}_{1}^{2}f \right) + \cdots,
\end{equation}
is defined in terms of the Lie-transform operator ${\sf T}_{n}^{-1} \equiv \exp(-\delta^{n}{\cal L}_{n})$, with the $n$th-order Lie derivative ${\cal L}_{n} \equiv {\sf G}_{n}\cdot\exd$ generated by the vector field ${\sf G}_{n}$, whose purpose is the remove the gyroangle dependence from the gyrocenter dynamics (i.e., the perturbed guiding-center dynamics). It is not entirely clear whether the existence of a non-perturbative integral expression for the gyrocenter magnetic moment \eqref{eq:mugy_integral} has any relevance for practical gyrokinetic numerical simulations. However, the statistical  full-orbit analysis used by Burby {\it et al.}\cite{j_w_burby_nonperturbative_2025} might still be able to yield some interesting results when applied to the perturbed guiding-center dynamics, which could be explored in future work.

In additional future work, we could explore whether the bounce-center dynamics\cite{Littlejohn_bc_1982,Cary_2009}, with the bounce-center action defined as
\begin{equation} 
    J_{\rm bc} \;\equiv\; \frac{1}{2\pi}\oint p_{\|}\,ds \;=\; \frac{1}{2\pi}\int_{0}^{2\pi}p_{\|}\,\pd{s}{\xi}\;d\xi, 
    \label{eq:bounce_action}
\end{equation}
satisfies its own Kruskal identity, where $\xi$ denotes the bounce angle. In this case, only one spatial symmetry needs to exist in order to yield conditions needed for proving the Kruskal identity. For this purpose, we will explore our previous work\cite{Duthoit_Brizard_2010} involving an axisymmetric dipole magnetic field. 

\appendix 

\section{\label{sec:screw_pinch}Kruskal Identity for the Slab Magnetic Field}

In this Appendix, we present the proof of the Kruskal identity \eqref{eq:Kruskal} for the normalized slab magnetic field ${\bf B}(y) = (1 + y)\,\wh{\sf z}$, which is summarized in the recent paper by Holas {\it et al.} \cite{Holas_etal_2025}, while a detailed analysis was previously presented elsewhere\cite{Brizard_2022}. The charged-particle orbits in a slab magnetic field are solutions of the dimensionless dynamical equations
\begin{eqnarray}
    \dot{x} &=& \epsilon\,v_{x}, \label{eq:x_dot} \\
    \dot{v}_{x} &=& v_{y}\,(1 + y), \label{eq:vx_dot} \\
    \dot{y} &=& \epsilon\,v_{y}, \label{eq:y_dot} \\
    \dot{v}_{y} &=& -\,v_{x}\,(1 + y), \label{eq:vy_dot}
\end{eqnarray}
where a dot refers to a time derivative with respect to the normalized time $(eB_{0}/m)\,t$ (only positive ions are considered here) and the normalized spatial coordinates $(x,y)$ denote the position of the charge particle. Since we are ignoring motion along the $z$-axis, we set the canonical momentum $P_{z} \equiv \dot{z} = 0$, so that the orbital motion remains on the $z = z_{0}$ surface.

The particle Lagrangian for this problem is expressed as
\begin{equation}
L \;=\;  -\,\left(y + \frac{1}{2}\,y^{2}\right)\;\dot{x} \;+\; \frac{1}{2} \left(\dot{x}^{2} \;+\; \dot{y}^{2}\right),
\label{eq:Lag_slab}
\end{equation}
where the magnetic field ${\bf B} = (1 + y)\,\wh{\sf z} \equiv \nabla\btimes{\bf A}$ is expressed in terms of the vector potential ${\bf A}(y) = -\,\left(y + \frac{1}{2}\,y^{2}\right)\,\wh{\sf x}$ and, because of the symmetry along the $x$-axis, the canonical momentum
\begin{equation}
    P_{x} \;=\; \pd{L}{\dot{x}} \;=\; \dot{x} - \left(y + \frac{1}{2}\,y^{2}\right) \;\equiv\; -\,\left(u - \frac{1}{2}\right)
    \label{eq:Px_def}
\end{equation}
is a dynamical invariant. If we now use the Routh reduction procedure, we obtain the reduced Lagrangian
\begin{equation}
L_{r}(y,\dot{y}) \;\equiv\; L \;-\; \dot{x}\;\pd{L}{\dot{x}} \;=\; \frac{1}{2}\,\dot{y}^{2} \;-\; V(y),
\label{eq:Lag_red_slab}
\end{equation}
where the reduced potential is
\begin{equation}
V(y) \;\equiv\; \frac{1}{2} \left[ \frac{1}{2}\,(1 + y)^{2} \;-\; u \right]^{2},
\label{eq:V_slab}
\end{equation}
which is a double-well quartic potential with minima $V(y_{\pm}) = 0$ at $y_{\pm} = -1 \pm\,\sqrt{2u}$ and a local maximum at $V(-1) = u^{2}/2$. If the energy $0 < E < u^{2}/2$, then the particle orbit is either trapped in the left well (-) or the right well (+).

In addition to the total energy constant of motion:
\begin{equation}
E \;=\; \frac{1}{2}\,\epsilon^{2}v_{\bot}^{2} \;=\; \frac{1}{2}\,\dot{y}^{2} \;+\; V(y),
\label{eq:E_slab}
\end{equation}
the reduced action integral
\begin{equation}
J_{r}(E,P_{x}) \;\equiv\; \frac{1}{2\pi}\,\oint P_{y}\;dy \;=\;  \frac{1}{2\pi}\,\oint \sqrt{2[E - V(y)]}\;dy,
\label{eq:Jr_slab} 
\end{equation}
is also an exact invariant, where $P_{y} \equiv \partial L/\partial\dot{y} = \dot{y} = \sqrt{2[E - V(y)]}$ is expressed in terms of the reduced quartic potential \eqref{eq:V_slab}. While this integral can be evaluated explicitly, we now turn our attention to the Newtonian gyromotion equations in order to find a simple expression for the reduced action integral \eqref{eq:Jr_slab}.

\subsection{Newtonian gyromotion equations}

If we now use the gyromotion representation 
\begin{equation}
    \left. \begin{array}{lcr}
    v_{x} &=& -\,v_{\bot}\,\sin\zeta \;=\; v_{\bot}\,\cos(2\varphi) \\
    v_{y} &=& -\,v_{\bot}\,\cos\zeta \;=\; -\;v_{\bot}\,\sin(2\varphi)
    \end{array} \right\},
    \label{eq:v_gyro}
\end{equation}
where $\zeta \equiv 2\,\varphi - \pi/2$ denotes the gyrorangle, we obtain the gyromotion equation 
\begin{equation}
\dot{\zeta} \;=\; 1 + y. 
\label{eq:zeta_y}
\end{equation}
Next, by using Eq.~\eqref{eq:y_dot}, we find $\dot{y} = -\,\epsilon\,v_{\bot}\,\cos\zeta$, so that we obtain
\[ \dot{y}^{2} \;=\; \epsilon^{2}\,v_{\bot}^{2}\,\cos^{2}\zeta \;=\; 2\,E \,-\, 2\,V(y) \;=\; \epsilon^{2}\,v_{\bot}^{2} \;-\; \left(\frac{1}{2}\,\dot{\zeta}^{2} \;-\; u \right)^{2}, \]
which yields
\begin{equation}
\dot{\varphi} \;=\; \nu_{0}\,\sqrt{(1 + {\sf e}) \;-\; 2\,{\sf e}\,\sin^{2}\varphi} \;\equiv\; \nu\;\sqrt{1 - m\,\sin^{2}\varphi},
\label{eq:varphi_dot}
\end{equation}
where ${\sf e} \equiv \epsilon\,v_{\bot}/u$, and  we introduced the definitions\cite{Brizard_2022}
\begin{equation}
    \nu \;\equiv\; \sqrt{\frac{u}{2}\,(1 + {\sf e})} \;\equiv\; \nu_{0}\,\sqrt{1 + {\sf e}}\;\;{\rm and}\;\; m \;\equiv\; \frac{2\,{\sf e}}{1 + {\sf e}}, 
    \label{eq:nu-m_def}
\end{equation} 
which are both constants of the motion. Using the gyromotion equation \eqref{eq:zeta_y}, with $\dot{\zeta} = 2\,\dot{\varphi}$, we obtain the solutions
\begin{equation}
    y_{\pm}(\varphi) \;=\; -\,1 \;\pm\; \sqrt{2u}\;\sqrt{(1 + {\sf e}) \;-\; 2\,{\sf e}\,\sin^{2}\varphi},
    \label{eq:y_varphi}
\end{equation}
 and the signs $\pm$ refer to the left well $(-)$ or the right well $(+)$.

We now focus our attention on motion in the right well (+), so that the solution of the differential equation \eqref{eq:varphi_dot} is expressed as
\begin{equation}
    \varphi(\tau) \;=\; {\rm arcsin}\left[
    {\rm sn}\left( \left.\nu\,\tau \;\right| m\right)\right],
    \label{eq:phi_sol}
\end{equation}
where we use the A\&S notation\cite{abramowitz2006handbook} ${\rm sn}(z|m)$ for the Jacobi ${\rm sn}$ function (see Chap. 16 by L.M. Milne-Thomson). At $\tau = 0$, the initial phase $\varphi(0) = 0$ implies that 
$y(0) = -1 + \sqrt{2u\,(1 + {\sf e})}$, i.e., the particle starts at the right turning point, where $v_{y}(0) = 0$, and after a half period $T/2$, the particle reaches the left turning point $y(T/2) = -1 + \sqrt{2u\,(1 - {\sf e})}$.

\subsection{Periodic $y$-motion}

We now proceed with the exact solutions for the motion of the charged particle in the $(x,y)$ plane, as it is trapped in the right well $(1 + y > 0)$. Using Eq.~(\ref{eq:phi_sol}) and the identity
\[ \sqrt{1 - m\,{\rm sn}^{2}(z|m)} \;=\; {\rm dn}(z|m), \]
we obtain the solution for $y(\tau) = -1 + 2\,\dot{\varphi}$ in the right well:
\begin{equation}
    y(\tau) \;=\; -\,1 \;+\; 2\nu\;\sqrt{1 - m\,\sin^{2}\varphi} 
    \;\equiv\; -\,1 \;+\; 2\,\nu\;{\rm dn}(\nu\,\tau|m).
    \label{eq:y_Jacobi}
\end{equation}
This solution is periodic with period 
\begin{equation}
    T \;\equiv\; 2\,{\sf K}(m)/\nu,
\end{equation}
where ${\sf K}(m) \equiv \int_{0}^{\pi/2} d\phi/\sqrt{1 - m\,\sin^{2}\phi}$ is the complete elliptic integral of the first kind (here, we use the A\&S notation\cite{abramowitz2006handbook} presented in Chap. 17 by L.M. Milne-Thomson).  Here, ${\rm dn}(0|m) = 1$ and
\[ {\sf dn}({\sf K}|m) \;=\; \sqrt{1 - m} \;=\; \sqrt{(1 - {\sf e})/(1 + {\sf e})}, \]
so that the range of $y(\tau)$ is
\[ -\,1 \;+\; \sqrt{2\,(u + \epsilon\,v_{\bot})} \;\leq\; y(\tau) \;\leq\; -\,1 
\;+\; \sqrt{2\,(u - \epsilon\,v_{\bot})}, \]
as expected for a particle trapped in the right well.

\subsection{Kruskal identity}

The reduced action integral \eqref{eq:Jr_slab} can now be expressed as
\begin{equation}
    J_{r} \;\equiv\; \frac{1}{2\pi} \oint \dot{y}\;dy \;=\; \frac{1}{2\pi} \oint \dot{y}^{2}\,d\tau \;=\; \frac{2}{\pi}\,m^{2}\nu^{3}\int_{0}^{2{\sf K}(m)} {\rm sn}^{2}(z|m)\;{\rm cn}^{2}(z|m)\; dz,
    \label{eq:J_def}
\end{equation}
where we used Eq.~(\ref{eq:y_Jacobi}) and the definition ${\rm dn}^{\prime}(z|m) \equiv -\,m\,{\rm sn}(z|m)\,{\rm cn}(z|m)$. Next, we solve the integral in terms of the complete elliptic integrals $({\sf E},{\sf K})$ as\cite{Brizard_2022}
\begin{equation}
    J_{r} \;\equiv\; \frac{2\,\nu^{3}}{3\pi} \left[ (4 - 2\,m)\,{\sf E}(m) \;-\; 4\,(1 - m)\,{\sf K}(m) \right].
    \label{eq:J_Fm}
\end{equation}
If we now expand Eq.~(\ref{eq:J_Fm}) in powers of ${\sf e} \ll 1$ up to fourth order, with $\nu^{3} = \nu_{0}^{3}\,\sqrt{(1 + {\sf e})^{3}}$, we find
\begin{equation}
    J_{r} \;\simeq\; \nu_{0}^{3}{\sf e}^{2} \left( 1 \;+\; \frac{3}{32}\,{\sf e}^{2}\right),
    \label{eq:J_final}
\end{equation}
where the third-order term is missing from this expansion.

We will now proceed with a comparison of the action integral (\ref{eq:J_def}), expanded up to fourth order in powers of ${\sf e}$ in Eq.~(\ref{eq:J_final}), with the guiding-center magnetic moment
\begin{equation} 
\mu \;=\; \mu_{0} \;+\; \mu_{1} \;+\; \mu_{2} + \cdots,
\label{eq:mu_gc}
\end{equation}
where each term $\mu_{n} = {\sf e}^{2}\,(\mu_{n0}\,{\sf e}^{n} + \mu_{n1}\,{\sf e}^{n+1} + \cdots)$ in the expansion (\ref{eq:mu_gc}) is further expanded in powers of ${\sf e}$ when the solution \eqref{eq:y_Jacobi} is inserted. In a slab magnetic field ${\bf B} = (1 + y)\,\wh{\sf z}$, the guiding-center magnetic moment 
(\ref{eq:mu_gc}) is expressed in terms of the components\cite{Holas_etal_2025}
\begin{eqnarray}
    \mu_{0} &=& \frac{\epsilon^{2}v_{\bot}^{2}}{2\,(1 + y)} \;=\; 
    \frac{2\,\nu_{0}^{4}\,{\sf e}^{2}}{(1 + y)} \;=\; \nu_{0}^{3}{\sf e}^{2} \left[ 1 \;-\; \frac{1}{2}\,
    {\sf e}\,\cos(2\varphi) \;+\; \frac{3}{8}\,{\sf e}^{2}\,\cos^{2}(2\varphi) + \cdots\right], \label{eq:mu_0} \\
    \mu_{1} &=& \mu_{0}\,\vb{\rho}_{0}\bdot\nabla\ln B \;=\; \frac{4\,\nu_{0}^{6}\,{\sf e}^{3}}{(1 + y)^{3}}\;\cos(2\varphi) \nonumber \\
    &=& \frac{1}{2}\,\nu_{0}^{3}{\sf e}^{3}\,\cos(2\varphi) \left[ 1 \;-\; \frac{3}{2}\,
    {\sf e}\,\cos(2\varphi) + \cdots \right], 
    \label{eq:mu_1} \\
    \mu_{2} &=& G_{2}^{\mu} \;+\; \frac{1}{2} \left[ -\,\vb{\rho}_{0}
    \bdot\nabla + \mu_{1}\,\pd{}{\mu_{0}} + \left(\pd{\vb{\rho}_{0}}{\zeta}\bdot\nabla \ln B\right)\,\pd{}{\zeta}\right]\mu_{1} \nonumber \\
    &=& \frac{3}{32}\,\nu_{0}^{3}{\sf e}^{4}\left[ 1 \;+\; 4\,\cos^{2}(2\varphi) \right] + \cdots,
   \label{eq:mu_2}
\end{eqnarray}
where the normalized gyroradius vector is
\begin{equation}
    \vb{\rho}_{0} \;=\; \sqrt{\frac{2\mu_{0}}{(1+ y)}}\,
    \left(\wh{\sf x} \;\cos\zeta \;-\; \wh{\sf y}\;\sin\zeta\right) \;=\; \frac{2\,\nu_{0}^{2}\,{\sf e}}{(1 + y)}  \left[\wh{\sf x} \;\sin(2\varphi) \;+\; \wh{\sf y}\;\cos(2\varphi) \right],
    \label{eq:rho_0}
\end{equation}
and we used the expansions
\begin{eqnarray}
    \frac{1}{1 + y} &=& \frac{1}{2\nu_{0}\,\sqrt{(1 + {\sf e}) - 2{\sf e}\,
    \sin^{2}\varphi}} \nonumber \\
    &=& \frac{1}{2\nu_{0}} \left( 1 \;-\; \frac{1}{2}\,
    {\sf e}\,\cos(2\varphi) \;+\; \frac{3}{8}\,{\sf e}^{2}\,\cos^{2}(2\varphi) \;+\; {\cal O}({\sf e}^{3}) \right), \\
     \frac{1}{(1 + y)^{3}} &=& \frac{1}{8\nu_{0}^{3}\,\sqrt{[(1 + {\sf e}) - 2{\sf e}\,\sin^{2}\varphi]^{3}}} \;=\; \frac{1}{8\nu_{0}^{3}} \left( 1 \;-\; \frac{3}{2}\,{\sf e}\,\cos(2\varphi) \;+\; {\cal O}({\sf e}^{2}) \right),
\end{eqnarray}
in Eqs.~\eqref{eq:mu_0} and \eqref{eq:mu_1}, respectively, so that we finally obtain
\begin{equation}
    \mu \;=\; \nu_{0}^{3}{\sf e}^{2} \left( 1 \;+\; \frac{3}{32}\,{\sf e}^{2}
    \right),
\end{equation}
which is exactly equal to the reduced action integral (\ref{eq:J_final}). Hence, we see that the gyroangle-dependent terms cancel out exactly in order to verify the Kruskal identity \eqref{eq:Kruskal}. We also see that, the third-order terms in the sum $\mu_{0} + \mu_{1}$ cancel each other out exactly, as is also seen in Eq.~\eqref{eq:mu_final} for the case of the screw-pinch magnetic geometry.

\section{\label{sec:gc}Guiding-center Dynamics in Screw-pinch Magnetic Geometry}

In this Appendix, we present the guiding-center theory \cite{Cary_2009} of charged particle motion in the screw-pinch magnetic field described in Sec.~\ref{sec:screw}. The results derived here are used in Secs.~\ref{sec:mu} and \ref{sec:action} to obtain the main result (\ref{eq:Jr_mu}) of this work, which confirms the identity of the magnetic moment (\ref{eq:mu_final}) and the reduced radial action integral (\ref{eq:Jr_rzeta}) for the 
screw-pinch magnetic geometry.

\subsection{Guiding-center equations of motion}

The guiding-center dynamics is expressed in terms of the guiding-center phase-space coordinates $({\bf X}_{\rm gc},P_{\|{\rm gc}},\mu_{\rm gc},
\zeta_{\rm gc})$ as the reduced guiding-center equations of motion
\begin{eqnarray}
    \frac{d{\bf X}_{\rm gc}}{dt} &=& \frac{P_{\|{\rm gc}}}{m}\,\frac{{\bf B}^{*}}{B_{\|}^{*}} \;+\; \frac{c\bhat}{e\,B_{\|}^{*}}\btimes\mu_{\rm gc} \nabla B, 
    \label{eq:xgc_dot} \\
    \frac{dP_{\|{\rm gc}}}{dt} &=& -\;\mu_{\rm gc}\,\nabla B\bdot\frac{{\bf B}^{*}}{B_{\|}^{*}}, \label{eq:pgc_dot}
\end{eqnarray}
where we have returned to using physical units and all spatial vectors and functions are evaluated at the guiding-center position ${\bf X}_{\rm gc}$. The guiding-center gyromotion equations, on the other hand, are
\begin{equation}
    \frac{d\zeta_{\rm gc}}{dt} \;=\; \Omega \;+\; {\bf R}\bdot
    \frac{d{\bf X}_{\rm gc}}{dt} \;\;{\rm and}\;\; 
    \frac{d\mu_{\rm gc}}{dt} \;\equiv\; 0,
    \label{eq:zetagc_dot}
\end{equation}
where the gyrogauge vector ${\bf R} \equiv \nabla\wh{\bot}\bdot\wh{\rho} = \tau\,\bhat - R_{\bot}\,\wh{\sf e}_{2}$ is given by Eq.~(\ref{eq:gyrogauge}). Its presence ensures that the guiding-center gyrofrequency $d\zeta_{\rm gc}/dt$ is gyrogauge invariant in the following sense. Considering the gyroangle transformation $\zeta_{\rm gc}^{\prime} \equiv \zeta_{\rm gc} + \chi({\bf X}_{\rm gc})$, which yields ${\bf R}^{\prime} = \nabla\wh{\bot}^{\prime} \bdot\wh{\rho}^{\prime} = {\bf R} + \nabla\chi$, so that ${\bf R}^{\prime}\bdot d{\bf X}_{\rm gc}/dt = {\bf R}\bdot d{\bf X}_{\rm gc}/dt + d\chi/dt$ and $d\zeta_{\rm gc}^{\prime}/dt = d\zeta_{\rm gc}/dt$. These equations are derived by Lie-transform perturbation method\cite{Littlejohn_1983}, where the guiding-center dynamics in the reduced phase space $({\bf X}_{\rm gc},P_{\|{\rm gc}})$ is asymptotically decoupled from the fast gyromotion $(\mu_{\rm gc},\zeta_{\rm gc})$.

In Eqs.~(\ref{eq:xgc_dot})-(\ref{eq:pgc_dot}), the {\it symplectic} magnetic field is expressed up to first order in magnetic nonuniformity as
\begin{equation}
    {\bf B}^{*} \;=\; {\bf B} \;+\; (c/e)\,P_{\|{\rm gc}}\,\nabla\btimes\bhat \;=\; B\;\left[ \left( 1 +
    \frac{P_{\|{\rm gc}}\,\tau_{m}}{m\Omega}\right)\,\bhat \;+\; \frac{P_{\|{\rm gc}}\,\kappa}{m\Omega}\,\wh{\sf e}_{2}\right], 
\end{equation}
which yields the ratio
\begin{equation}
    \frac{{\bf B}^{*}}{B_{\|}^{*}} \;=\; \bhat \;+\; P_{\|{\rm gc}}\,\left(\frac{c\,\kappa}{e\,B_{\|}^{*}}\right)\,\wh{\sf e}_{2},
    \label{eq:Bstar}
\end{equation}
where $B_{\|}^{*} \equiv \bhat\bdot{\bf B}^{*} = B + P_{\|{\rm gc}}\,(c\tau_{m}/eB)$ is the guiding-center Jacobian. Here, we note that Eq.~(\ref{eq:Bstar}) does not have a radial component, which will play an important role below in establishing constants of motion.

\subsection{Guiding-center constants of motion}

First, since $\nabla B = -\,B\;(\Theta^{\prime}\,\tan\,\Theta)\;\wh{\sf e}_{1}$, we see that the guiding-center radial position $r_{\rm gc}$ is an exact guiding-center invariant:
\[ \frac{dr_{\rm gc}}{dt} \;\equiv\; \nabla r_{\rm gc}\bdot\frac{d{\bf X}_{\rm gc}}{dt} \;=\; -\,
\wh{\sf e}_{1}\bdot\frac{d{\bf X}_{\rm gc}}{dt} \;=\; 0, \] 
since the guiding-center velocity (\ref{eq:xgc_dot}) has no radial component:
\begin{equation}
    \frac{d{\bf X}_{\rm gc}}{dt} \;=\; \frac{P_{\|{\rm gc}}}{m}\,\bhat \;+\; \left(
    \frac{P_{\|{\rm gc}}^{2}\,\kappa}{m} \;-\; \mu_{\rm gc}\,B\;(\Theta^{\prime}\,\tan\,\Theta)\right)\,\frac{c\,\wh{\sf e}_{2}}{e\,B_{\|}^{*}},
\end{equation}
i.e., the guiding-center velocity has no radial component, which means that each guiding-center orbit lies on a constant cylindrical surface with guiding-center radius $r_{\rm gc}$. Next, Eq.~(\ref{eq:pgc_dot}) immediately leads to the conservation law of guiding-center parallel momentum: 
\begin{equation} 
    \frac{dP_{\|{\rm gc}}}{dt} \;=\;  -\;\mu_{\rm gc}\,\nabla B\bdot\frac{{\bf B}^{*}}{B_{\|}^{*}} \;=\; \mu_{\rm gc}\,B\;(\Theta^{\prime}\,
    \tan\Theta)\;\wh{\sf e}_{1}\bdot\frac{{\bf B}^{*}}{B_{\|}^{*}} \;=\; 0, 
\end{equation}
since $\wh{\sf e}_{1}\bdot{\bf B}^{*} = 0$. Hence, the guiding-center pitch angle $\lambda_{\rm gc} \;\equiv \cos^{-1}(P_{\|{\rm gc}}/P_{\rm gc})$ is also a guiding-center constant of motion. 

\subsection{Guiding-center transformation in screw-pinch magnetic field}

Now that we have established that the radial guiding-center position $r_{\rm gc}$ and the guiding-center pitch angle $\lambda_{\rm gc}$ are constants of motion in screw-pinch magnetic geometry, we will now use the inverse guiding-center transformation \cite{Littlejohn_1983,tronko_lagrangian_2015}
\begin{equation}
    z^{\alpha} \;=\; Z_{\rm gc}^{\alpha} \;-\; \epsilon\,G_{1}^{\alpha} \;-\; \epsilon^{2} \left( G_{2}^{\alpha} \;-\; \frac{1}{2}\,{\sf G}_{1}\cdot\exd 
    G_{1}^{\alpha}\right) \;+\; \cdots,
\end{equation}
where the phase-space vector-field components $G_{n}^{\alpha}$ generate the transformation (or its inverse) at $n^{th}$ order.

\subsubsection{Radial position}

First, the particle position ${\bf x}$ and the guiding-center position ${\bf X}_{\rm gc}$ are defined in terms of the expansion
\begin{equation}
    {\bf x} \;=\; {\bf X}_{\rm gc} \;+\; \vb{\rho}_{0} \;+\; \cdots,
\end{equation}
where the lowest-order guiding-center gyroradius $\vb{\rho}_{0}$ is given by Eq.~(\ref{eq:rho_0}). The radial position $r \equiv r_{\rm gc} + \vb{\rho}_{0}\bdot\nabla r + \cdots$ is, therefore, expressed up to first order in $\epsilon$ as
\begin{equation}
    r \;=\; r_{\rm gc} \;+\; \wh{\sf r}\bdot\vb{\rho}_{0} \;=\; r_{\rm gc} \;-\; \wh{\sf e}_{1}\bdot\vb{\rho}_{0} \;=\; r_{\rm gc} \;-\; \epsilon\,\rho_{\bot}\,\cos\zeta \;\equiv\; r_{0} \;+\; \epsilon\,r_{1},
\end{equation}
where the guiding-center radial position $r_{\rm gc}$ is a guiding-center constant of motion. Hence, we find the lowest-order radial position $r_{0} = r_{\rm gc}$ and the first-order radial position is defined as
\begin{equation} 
    r_{1} \;\equiv\; -\;\rho_{\bot}\,\cos\zeta \;=\; -\;v_{\bot}\,\cos\Theta\;\cos\zeta.
    \label{eq:r1gc}
\end{equation}
We immediately note that
\begin{equation}
    \frac{dr}{dt} \;=\; \frac{dr_{0}}{dt} \;+\; \epsilon\,\frac{dr_{1}}{dt} \;=\; \epsilon\,v_{\bot}\,\cos\Theta\;\sin\zeta\;\frac{d\zeta}{dt} \;+\; \cdots \;=\; \epsilon\;v_{\bot}\,\sin\zeta \;+\; {\cal O}(\epsilon^{2}),
\end{equation}
where we used the lowest-order gyrofrequency $d\zeta/dt = \sec\Theta + \cdots$ and the guiding-center radial position $r_{0}$ is also a constant of particle motion (up to first order in $\epsilon$).

\subsubsection{Pitch angle}

Next, using the pitch-angle identity $p_{\|} \equiv p\,\cos\lambda$, we find the first-order pitch-angle generator $G_{1}^{\lambda} = -\,G_{1}^{p_{\|}}/p_{\bot} \equiv -\,\lambda_{1}$ expressed in terms of the parallel momentum generator\cite{Littlejohn_1983} $G_{1}^{p_{\|}}$. According to standard guiding-center theory \cite{Littlejohn_1983,tronko_lagrangian_2015}, the lowest-order pitch angle $\lambda_{0} \equiv \lambda_{\rm gc}$ is, therefore, the guiding-center pitch angle, and the first-order correction is
\begin{equation}
    \lambda_{1} \;\equiv\; \frac{1}{2}\,v_{\bot}\cos\Theta\;\left(\tau_{m} + \alpha_{1}\right) \;-\; v_{\|}\,\cos\Theta\;\kappa\;\cos\zeta,
\end{equation}
where $(\tau_{m} + \alpha_{1})$ is given by Eq.~(\ref{eq:tau_alpha}). Once again, we immediately note that
\begin{eqnarray} 
    \frac{d\lambda}{dt} &=& \frac{d\lambda_{0}}{dt} \;+\; \epsilon\,\frac{d\lambda_{1}}{dt} \;=\; \epsilon\,\sin\zeta \left[ v_{\|}\,\kappa \;+\;
    v_{\bot}\,\cos\zeta\,\left(\tau \;-\; \Theta^{\prime}\right) \right] \cos\Theta\;\frac{d\zeta}{dt} \;+\; \cdots \nonumber \\
    &=& \epsilon\,\sin\zeta \left[ v_{\|}\,\kappa \;+\; v_{\bot}\,\cos\zeta\,\left(\tau \;-\; \Theta^{\prime}\right) \right] \;+\; {\cal O}(\epsilon^{2}),
\end{eqnarray}
where we used the lowest-order gyrofrequency $d\zeta/dt = \sec\Theta + \cdots$ and the guiding-center pitch angle $\lambda_{0}$ is also a constant of particle motion (up to first order in $\epsilon$).

\subsection{Guiding-center magnetic-moment action integral}

In the Introduction, we stated that we could obtain the lowest-order magnetic moment from the guiding-center magnetic-moment action integral (\ref{eq:mugc_integral}):
\begin{equation}
    \mu_{\rm gc} \;=\; \frac{1}{2\pi}\int_{0}^{2\pi} \left(\epsilon\,{\bf v} \;+\; \psi\;\nabla\theta \;-\; \Psi\;\nabla z\right)\bdot\epsilon\,
    \pd{\vb{\rho}_{0}}{\zeta}\;d\zeta.
\end{equation}
First, using $\vb{\rho}_{0} = v_{\bot}\,\cos\Theta\;\wh{\bot}$, we obtain
\begin{equation}
    \mu_{\rm gc} \;=\; \frac{1}{2\pi}\int_{0}^{2\pi} \left[ \epsilon^{2}\,v_{\bot}^{2}\,\cos\Theta \;+\; \epsilon\,v_{\bot}\,\cos\Theta\;\cos\zeta\left( \psi\;\frac{\cos\Theta}{r} + \Psi\,\sin\Theta\right) \right]d\zeta.
    \label{eq:mugc_action}
\end{equation}
Next, using Eqs.~(\ref{eq:psi_ovpsi})-(\ref{eq:Psi_ovPsi}), where $(\ov{\psi} = P_{\theta}, \ov{\Psi} = -\,P_{z})$ represent the invariant canonical momenta, we obtain
\[ \psi\;\frac{\cos\Theta}{r} + \Psi\,\sin\Theta \;=\; \ov{\psi}\;\frac{\cos\Theta}{r} + \ov{\Psi}\,\sin\Theta \;-\; \epsilon\,v_{\bot}\,\cos\zeta, \]
so that Eq.~(\ref{eq:mugc_action}) becomes
\begin{equation}
    \mu_{\rm gc} \;=\; \frac{1}{2\pi}\int_{0}^{2\pi} \left[ \epsilon^{2}\,v_{\bot}^{2}\,\cos\Theta\,\sin^{2}\zeta \;+\; \epsilon\,v_{\bot}\,\cos\Theta\;\cos\zeta\left( \ov{\psi}\;\frac{\cos\Theta}{r} + \ov{\Psi}\,\sin\Theta\right) \right]d\zeta.
    \label{eq:mugc_final}
\end{equation}
Hence, to lowest order in $\epsilon$, we find $\mu_{\rm gc} = \frac{1}{2}\,\epsilon^{2}\,v_{\bot0}^{2}\,\cos\Theta_{0}$.

\bibliographystyle{unsrt}
\bibliography{screw_pinch.bib}

@PREAMBLE{
 "\providecommand{\noopsort}[1]{}" 
 # "\providecommand{\singleletter}[1]{#1}%" 
}

@BOOK{Northrop_1963,
   author       = {T. G. Northrop},
   lccn         = {lc63022462},
   series       = {Interscience tracts on physics and astronomy},
   year         = 1963,
   title        = {The Adiabatic Motion of Charged Particles},
   publisher    = {Interscience Publishers}
}

@BOOK{Goldstein_2002,
   author       = {H. Goldstein and C. Poole and J. Safko},
   year         = 2002,
   title        = {Classical Mechanics, 3rd ed.},
   publisher    = {Addison-Wesley}
}

@ARTICLE{Brizard_2007,
  title = {Foundations of nonlinear gyrokinetic theory},
  author = {Brizard, A. J. and Hahm, T. S.},
  journal = {Rev. Mod. Phys.},
  volume = {79},
  issue = {2},
  pages = {421--468},
  year = {2007},
  month = {Apr},
  doi = {10.1103/RevModPhys.79.421},
  publisher = {American Physical Society}
}

@ARTICLE{Cary_2009,
   author       = "J. R. Cary and A. J. Brizard",
   title           = "Hamiltonian theory of guiding-center motion",
   year         = "2009",
   journal      = "Rev. Mod. Phys.",
   volume       = "81",
   pages        = "693"
}

@ARTICLE{Kruskal_1962,
   author       = "M. Kruskal",
   title 				= "Asymptotic Theory of Hamiltonian and other Systems with all Solutions nearly periodic",
   year         = "1962",
   journal      = "J. Math. Phys.",
   volume       = "3",
   pages        = "806"
}

@ARTICLE{Littlejohn_1983,
   author       = "R. G. Littlejohn",
   title 				= "Variational principles of guiding centre motion",
   year         = "1983",
   journal      = "J. Plasma Phys.",
   volume       = "29",
   pages        = "111"
}

@MISC{Burby_loops_2019,
   author   ="J. W. Burby",
   title = "Guiding center dynamics as motion on a formal slow manifold in loop space",
   eprint    ="arXiv:1905.04410",
   year     ="2019"
}

@article{j_w_burby_nonperturbative_2025,
    title = {Nonperturbative guiding center model for magnetized plasmas},
    volume = {134},
    journal = {Phys. Rev. Lett.},
    author = {{J. W. Burby} and {I. A. Maldonado} and {M. Ruth} and {D. A. Messenger} and {L. Carbajal}},
    year = {2025},
    pages = {175101},
}

@article{Brizard_2022,
    title = {Action–angle coordinates for motion in a straight magnetic field with constant gradient},
    volume = {114},
    issn = {10075704},
    url = {https://linkinghub.elsevier.com/retrieve/pii/S1007570422002325},
    doi = {10.1016/j.cnsns.2022.106652},
    urldate = {2025-07-03},
    journal = {Comm. Nonlin. Sci. Num. Sim.},
    author = {Brizard, A. J.},
    month = nov,
    year = {2022},
    pages = {106652},
}

@article{burby_automation_2013,
    title = {Automation of the guiding center expansion},
    volume = {20},
    issn = {1070-664X, 1089-7674},
    url = {https://pubs.aip.org/pop/article/20/7/072105/107086/Automation-of-the-guiding-center-expansion},
    doi = {10.1063/1.4813247},
    number = {7},
    urldate = {2025-07-11},
    journal = {Phys. Plasmas},
    author = {Burby, J. W. and Squire, J. and Qin, H.},
    month = jul,
    year = {2013},
    note = {Publisher: AIP Publishing},
}

@article{tronko_lagrangian_2015,
    title = {Lagrangian and {Hamiltonian} constraints for guiding-center {Hamiltonian} theories},
    volume = {22},
    issn = {1070-664X, 1089-7674},
    url = {https://pubs.aip.org/pop/article/22/11/112507/108546/Lagrangian-and-Hamiltonian-constraints-for-guiding},
    doi = {10.1063/1.4935925},
    number = {11},
    urldate = {2025-08-11},
    journal = {Phys. Plasmas},
    author = {Tronko, N. and Brizard, A. J.},
    month = nov,
    year = {2015},
    pages = {112507},
}

@article{burby_nearly_2023,
    title = {Nearly {Periodic} {Maps} and {Geometric} {Integration} of {Noncanonical} {Hamiltonian} {Systems}},
    volume = {33},
    issn = {1432-1467},
    url = {https://doi.org/10.1007/s00332-023-09891-4},
    doi = {10.1007/s00332-023-09891-4},
    number = {2},
    urldate = {2025-06-26},
    journal = {J. Nonlin. Sci.},
    author = {Burby, J. W. and Hirvijoki, E. and Leok, M.},
    month = feb,
    year = {2023},
    pages = {38},
}

@article{Littlejohn_bc_1982,
  title={Hamiltonian theory of guiding center bounce motion},
  author={Littlejohn, R. G.},
  journal={Physica Scripta},
  volume={1982},
  number={T2A},
  pages={119},
  year={1982},
  publisher={IOP Publishing}
}

@article{Holas_etal_2025,
    title = {Exact expressions for nonperturbative guiding center theory in symmetric fields},
   author = {I. Holas and R. Agarwal and J. W. Burby and A. J. Brizard},
  journal = {arXiv:2508.09484},
     year = {2025}
}

@book{Brizard:2015,
  title={An Introduction to Lagrangian Mechanics (2nd ed.)},
  author={Brizard, A. J.},
  volume={},
  year={2015},
  publisher={World Scientific}
}

@article{Littlejohn_1982,
  title={Hamiltonian perturbation theory in noncanonical coordinates},
  author={Littlejohn, R. G.},
  journal={J. Math. Phys.},
  volume={23},
  number={5},
  pages={742--747},
  year={1982},
  publisher={American Institute of Physics}
}

@article{Brizard_2017,
    author = {Brizard, A. J.},
    title = {On the validity of the guiding-center approximation in the presence of strong magnetic gradients},
    journal = {Phys. Plasmas},
    volume = {24},
    pages = {042115},
    year = {2017},
    }

@article{Brizard_Markowski_2022,
    author = {Brizard, A. J. and Markowski, D. G.},
    title = {On the validity of the guiding-center approximation in a magnetic dipole field},
    journal = {Phys. Plasmas},
    volume = {29},
     pages = {022101},
    year = {2022}
}

@article{Northrop_Liu_Kruskal_1966,
    author = {Northrop, T. G. and Liu, C. S. and Kruskal, M. D.},
    title = {First Correction to the Second Adiabatic Invariant of Charged‐Particle Motion},
    journal = {Phys. Fluids},
    volume = {9},
     pages = {1503},
    year = 1966}

@article{Northrop_Teller_1960,
  title = {Stability of the Adiabatic Motion of Charged Particles in the Earth's Field},
  author = {Northrop, T. G. and Teller, E.},
  journal = {Phys. Rev.},
  volume = {117},
  issue = {1},
  pages = {215--225},
  year = {1960}
  }

@article{Duthoit_Brizard_2010,
    author = {Duthoit, F.-X. and Brizard, A. J. and Peysson, Y. and Decker, J.},
    title = {Perturbation analysis of trapped-particle dynamics in axisymmetric dipole geometry},
    journal = {Phys. Plasmas},
    volume = {17},
    number = {10},
    pages = {102903},
    year = {2010}
}

@article{Tao_Chan_Brizard_2007,
    author = {Tao, X. and Chan, A. A. and Brizard, A. J.},
    title = {Hamiltonian theory of adiabatic motion of relativistic charged particles},
    journal = {Phys. Plasmas},
    volume = {14},
    number = {9},
    pages = {092107},
    year = {2007}
}

@ARTICLE{Brizard_2023,
   author       = "A. J. Brizard and B. C. Hodgeman",
   title        = "Faithful guiding-center orbits in an axisymmetric magnetic
field",
   year         = "2023",
   journal      = "Phys. Plasmas",
   volume       = "30",
   pages        = "042115",
   doi = "http://dx.doi.org/10.1063/5.0145035"
}

@ARTICLE{Belova_2003,
   author       = "E. V. Belova and N. N. Gorelenkov and C. Z. Cheng",
   title        = "Self-consistent equilibrium model of low aspect-ratio toroidal plasma with
energetic beam ions",
   year         = "2003",
   journal      = "Phys. Plasmas",
   volume       = "10",
   pages        = "3240",
   doi = "http://link.aip.org/link/doi/10.1063/1.1592155?ver=pdfcov"
}

@book{abramowitz2006handbook,
  title={Handbook of mathematical functions: with formulas, graphs, and mathematical tables},
  author={Abramowitz, M. and Stegun, I. A.},
  year={2006},
  publisher={National Bureau of Standards Washington, DC}
}

\end{document}